\newcommand{\gd}{\dot{\gamma}} 
\newcommand{\hop}{\omega}      
\newcommand{\yield}{\sigma_{\rm Y}} 
\newcommand{\eff}{\Delta E}    

\documentstyle[multicol,aps,psfig]{revtex}
\begin{document}

\draft

\title{Jamming, hysteresis and oscillation
in scalar models for shear thickening}

\author{D. A. Head,$^{1}$ A. Ajdari,$^{2}$ and M. E. Cates$^{1}$}

\address{$^{1}$Department of Physics and Astronomy, JCMB King's Buildings,
University of Edinburgh, Edinburgh EH9 3JZ, UK\\
$^{2}$Laboratoire de Physico--Chimie Th\'eorique, Esa CNRS 7083,
ESPCI, 10 rue Vauquelin, F-75231 Paris Cedex 05, France}

\date{\today}

\maketitle

\begin{abstract}
We investigate shear thickening and jamming within the framework
of a family of spatially homogeneous,
scalar rheological models.
These are based on the `soft glassy rheology'
model of Sollich {\em et al.}
[Phys. Rev. Lett.~{\bf 78}, 2020 (1997)],
but with an effective temperature $x$ that is a
decreasing function of
either the global stress $\sigma$ or the local strain~$l$.
For appropiate $x=x(\sigma)$, it is shown that the flow curves
include a region of negative slope,
around which the stress exhibits hysteresis under a
cyclically varying imposed strain rate~$\gd$.
A subclass of these $x(\sigma)$ have flow curves that
touch the $\gd=0$ axis for a finite range of stresses;
imposing a stress from this range {\em jams} the system,
in the sense that the strain $\gamma$ creeps only logarithmically
with time~$t$, $\gamma(t)\sim\ln t$.
These same systems may produce
a finite asymptotic yield stress under an imposed strain,
in a manner that depends
on the entire stress history of the sample,
a phenomenon we refer to as {\em history--dependent jamming}.
In contrast, when $x=x(l)$ the flow curves are always monotonic,
but we show that some $x(l)$ generate an oscillatory strain response
for a range of steady imposed stresses.
Similar spontaneous oscillations are observed in a simplified model
with fewer degrees of freedom.
We discuss this result in relation to the temporal instabilities
observed in rheological experiments
and stick--slip behaviour found in other contexts,
and comment on the possible relationship with
`delay differential equations'
that are known to produce oscillations and chaos.
\end{abstract}


\maketitle
\begin{multicols}{2}
\narrowtext

\section{Introduction}
\label{s:intro}

A wide variety of materials can be driven into a
non--equilibrium state that is either solid--like
and static, or fluid--like but only relaxes on
time scales that far exceed the experimental time frame,
if at all.
Such states are often referred to as `jammed,'
and are realisable in molecular liquids having undergone
a rapid quench to low temperatures, or colloids at a high
volume fraction, to cite just two examples~\cite{glass,jam_sam}.
It has recently been postulated by Liu and Nagel that the
nature of the jammed state may be independent of
the manner in which it was formed~\cite{jam_phase}.
For example, a stress--induced jamming transition may produce
a qualitatively similar state to a temperature--induced transition.
This was expressed in the form of a `jamming phase diagram,'
in which jammed configurations occupy a compact region
near to the origin of a phase space comprised of three axes~:
the temperature~$T$, the volume~$V$ and
the load~$\sigma$~\cite{jam_phase,jam_attr}.

In its simplest form, the Liu--Nagel jamming phase diagram
suggests that increasing the applied load can only
increase the likelihood of flow.
However, it is equally feasible for an applied
load to induce a jammed state.
For instance, if a pile of sand is formed and then
gravity is switched off, the system unjams without
any significant variation in volume.
Thus the the load (here controlled by gravity) jams the system.
This concurs with the earlier claim that jammed
systems may be classified as `fragile'
--- that is, they can support only certain, {\em compatible} loads,
and will rearrange or flow under an incompatible load~\cite{fragile}.
Furthermore, one could also envisage a class of loadings
that can alternately induce and destroy a jammed state
as the magnitude of the load increases,
a situation that could be referred to as
`re-entrant jamming.'

The purpose of this paper is to investigate the
nature of transitions to or from jammed configurations
that have, as their control parameter,
the imposed shear stress~$\sigma$.
We do this by providing concrete examples
of models that exhibit jamming transitions with~$\sigma$,
including instances of the re-entrant jamming scenario described above.
These models are based on
the `soft glassy rheology' (SGR) model of
Sollich {\em et al.}~\cite{sgr_prl,sgr_pre,sgr_aging},
which was originally devised to highlight
the possible existence of glassy relaxation in a range
of soft materials, such as foams, emulsions, pastes~{\em etc.}
It is parameterised by an effective temperature~$x$,
which in in the context of soft matter is thought to
represent some form of mechanical noise,
but may refer to the true thermal temperature in other applications.

As it was originally defined, the SGR model only exhibits shear thinning,
which is clearly unsuitable for our purposes.
Therefore we consider variants of the model in which
the effective temperature $x$ is no longer constant,
but can vary with the state of the system.
More precisely, $x$ is treated a function of both the global
stress~$\sigma$ and the local strain~$l$, {\em i.e.}~$x=x(\sigma,l)$.
By choosing suitable forms of~$x(\sigma,l)$,
systems can be constructed that become `colder' as they become
more stressed,
which allows for the system to shear thicken and even `jam.'

For clarity, we restrict our attention to two
limiting forms of $x(\sigma,l)$, namely
$x(\sigma)$ and $x(l)$.
In the former case,
certain classes of $x(\sigma)$ are shown to exhibit a
flow curve
({\em i.e.} the curve of the stress $\sigma$ versus the strain rate $\gd$
under conditions of steady flow)
that is non--monotonic,
and can touch the $\gd=0$ axis for range of finite values of~$\sigma$.
Thus a `jammed' state with $\gd=0^{+}$ can be reached
for some $\sigma$ but not for others.
Furthermore, it is also shown that, for systems driven by
an imposed $\gamma(t)$, whether or not a jammed configuration
is approached at late times
depends on the entire strain history of the system,
and not just the behaviour of $\gamma(t)$ as $t\to\infty$.
We refer to this phenomenon as `history--dependent jamming.'

The behaviour of systems with $x=x(l)$ is somewhat different,
but no less remarkable.
When driven by an imposed~$\sigma$,
certain forms of $x(l)$ exhibit a regime
in which the viscosity never reaches its steady flow value
but oscillates in time, with a waveform
that is approximately sinusoidal near to the transition point,
but becomes increasingly sharp deep into the oscillatory regime.
The models considered here are spatially homogeneous,
and so this oscillation is purely temporal, having no spatial
component.
It is tempting to associate these oscillations
with the stick--slip
behaviour observed in granular systems and plate
tectonics~\cite{earthquake,thinliquids,granstickslip1,granstickslip2},
but we shall give reasons why we
believe that the underlying physics might be rather different.
We cannot rule out the possibility of more complex
oscillatory behaviour, maybe even chaotic,
arising in as--of--yet unobserved regions of parameter space.

The finding of a bifurcation to oscillatory behaviour
for $x=x(l)$ is all the more remarkable because the flow curve
(as already defined)
is everywhere monotonic increasing for this class of~$x$.
By contrast, instances of rheological instabilities that have been found
experimentally have tended to occur for ranges of parameters in which the
flow curve has a negative
gradient~\cite{olmsted,lu,volkova,bonn,jacques,hu,laun}.
This suggests that mechanism behind the instability observed here
may be qualitatively different to those cited above,
and may be realisable in some range of materials
that has yet to be identified.

This paper is arranged as follows.
In Sec.~\ref{s:general} the class of models to be studied
is defined, with particular attention being paid to those
aspects that differ from the SGR model.
The known results of the SGR model that are relevant
to the subsequent discussion are then briefly summarised
in Sec.~\ref{s:sgr}.
Systems with $x=x(\sigma)$ are described in Sec.~\ref{s:global},
where it explained how the flow curves can be graphically
interpreted as mappings from the SGR flow curves.
The time--dependent behaviour of the system has
been found by numerical integration of the governing master equation.
An example of history--dependent jamming is presented,
and explained in terms of the stability of the steady flow
solutions.

In Sec.~\ref{s:local} we turn to consider the case $x=x(l)$,
and analytically prove that the flow curve is everywhere
monotonic.
Nonetheless, the simulations results presented here
clearly show that $\gd(t)$ can oscillate in time for a range
of imposed stresses~$\sigma$.
A qualitative explanation of the emergence of the oscillatory
phase is also given,
in terms of the model's internal degrees of freedom.
The intensive nature of the simulations has meant that
only a small range of functional forms for $x(l)$ have been investigated,
and hence it has not been possible to fully characterise
this range of models.
Therefore we consider a simplified model in Sec.~\ref{s:monodisperse},
for which the simulation times are significantly shorter
and a more complete picture has been realised.
Some results of this model have also been
presented elsewhere~\cite{EPL}.
This reduced model clearly shows that the mean value of $\gd$ during the
oscillatory motion deviates from the steady flow value by
as much as an order of magnitude.
Finally, in Sec.~\ref{s:disc} we discuss some outstanding issues
raised by this work, before summarising our results
in Sec.~\ref{s:summ}.

\section{Models of the SGR--type}
\label{s:general}

All of the models studied in this paper represent generalisations
of the SGR model of Sollich {\em et al.}~\cite{sgr_prl,sgr_pre,sgr_aging}.
It is therefore prudent to first describe those features
that are common to this class of models, before considering
each of the different generalisations in turn.
This is the purpose of the current section.
Since the various assumptions that lie behind
the construction of the SGR model have already
been discussed at length elsewhere,
we shall here give just a brief overview of the derivation
and refer the reader to~\cite{sgr_prl,sgr_pre,sgr_aging}
for more detailed physical arguments.
Only those aspects that differ from the SGR model
will be discussed in full.

Our goal is to construct deliberately simplified models
that exhibit shear--thickening and jamming,
but whose interpretation is more transparent than
that of a detailed microscopic model.
In this spirit,
the models in this class all
share a number of simplifications.
Only a single shear component of the strain and
stress tensors are considered, which will be denoted
by $\gamma$ and $\sigma$ respectively.
This is therefore a class of {\em scalar} models.

It is further assumed that the system can be coarse grained
into a collection of mesoscopic subsystems,
each of which are fully described by two scalar variables,
namely a local strain $l$ and a stability parameter~$E$.
These mesoscopic subsystems will be referred to as
{\em elements}.
We simply assume here that such a coarse--graining is possible,
notwithstanding the significant practical challenges
in constructing a suitable scheme for any given microscopic model.

At any given instant, each element has a probability of yielding
per unit time that is denoted by~$\Gamma=\Gamma(E,l)$.
When an element yields, it is assumed that its microscopic
constituents are rearranged to such a degree that it loses
all memory of its former configuration.
Its strain returns to zero, and it is assigned a new value
of $E$ that is drawn from a {\em prior} distribution
$\rho(E)$.
Suitable functional forms for
$\Gamma(E,l)$ and $\rho(E)$ will be discussed below.
The value of $E$ remains fixed until the element yields again,
but $l$ follows the global strain $\gamma$
according to $\dot{l}=\gd$.
Thus the elements are affinely deformed
by the flow field, which is assumed to be homogeneous.

Let us define $P(E,l,t)$ to be the probability density function
of elements which have a local strain $l$ and a barrier $E$
at time~$t$.
Then $P(E,l,t)$ evolves in time according to two distinct mechanisms~:
the homogeneous shearing at a rate~$\gd$,
and the yielding of elements at a rate
of $\Gamma(E,l)$ per unit time.
Thus the master equation for $P(E,l,t)$ is

\begin{eqnarray}
\frac{\partial}{\partial t}P(E,l,t)+
\gd
\frac{\partial}{\partial l}P(E,l,t)&=&
\mbox{}-\Gamma(E,l)P(E,l,t)
\nonumber\\
&&\mbox{}+
\hop(t)\,\delta(l)\,\rho(E)\quad.
\label{e:master}
\end{eqnarray}

\noindent{}The second term on the left--hand side
represents the increase in local strains $l$ according to
the uniform global strain rate~$\gd$.
The right hand side describes the yielding of elements.
The first term, which is negative, accounts for the
loss of elements as they yield at a rate $\Gamma(E,l)$.
Conversely, the second term represents these same elements
{\em after} they have yielded,
which have a strain $l=0$ and a value of $E$ drawn from $\rho(E)$.
The total rate of yielding $\hop(t)$ is defined by

\begin{eqnarray}
\hop(t)&=&
\int_{0}^{\infty}{\rm d}E\int_{-\infty}^{\infty}{\rm d}l\,
\Gamma(E,l)P(E,l,t)
\\
&=&
\langle\,\Gamma(E,l)\,\rangle\quad.
\label{e:yield}
\end{eqnarray}

\noindent{}Here we
have introduced the notation that the angled brackets
`$\langle\ldots\rangle$' represents the instantaneous
average of the given function over $P(E,l,t)$.
This will be used frequently in what follows.

The master equation (\ref{e:master}) only describes the
evolution of the strain degrees of freedom.
To characterise the rheological response of the system,
some relation must be found between the local
strains $l$ and the global stress~$\sigma$.
This involves two further assumptions.
Firstly, the elements are supposed to behave elastically between yield events,
so that the local stress is just~$kl$,
where $k>0$ is a uniform elastic constant.
Secondly, $\sigma$ is the arithmetic mean of the local stresses,
or $\sigma=\langle kl\rangle=k\langle l\rangle$.
Other averaging procedures could be employed~\cite{averaging},
but we focus here on the simplest non--trivial option.
Although the local stress--strain relationship is elastic,
the global stress also incorporates the yielding of elements
and, as will be seen below, the $\sigma$--$\gamma$ relationship
is not a linear one.

We have been unable to find an analytic solution to
the master equation~(\ref{e:master}) for any non--trivial~$\Gamma(E,l)$.
Instead it has been numerically integrated using the procedure
summarised in Appendix~\ref{s:numerics}.
However, it will sometimes be necessary to refer to the
steady state solution for a constant~$\gd\neq0$,
which can be found exactly,

\begin{equation}
P_{\infty}(E,l)
=
\frac{\hop_{\infty}}{\gd}\rho(E)\exp
\left[
-\frac{1}{\gd}
\int_{0}^{l}{\rm d}l'\,\Gamma(E,l')
\right]\quad.
\label{e:st3}
\end{equation}

\noindent{}This is derived by setting $\partial_{t}P=0$ in
(\ref{e:master}) and integrating the resulting first--order
ordinary differential equation with respect to~$l$.
The asymptotic yield rate $\hop_{\infty}\equiv\lim_{t\to\infty}\hop(t)$
can be found by requiring that the integral of
$P_{\infty}(E,l)$ is unity.

\subsection{Yielding modelled as an activated process}
\label{s:choice_of_gamma}

In the SGR model, the yield rate $\Gamma(E,l)$
was assigned a functional form similar to that of
an activated process~\cite{sgr_pre}.
This was based on the idea that each
element can be represented by a single particle moving on a
free energy landscape, which remains 
confined within a well of depth $E$
until random fluctuations allow it to cross over this
barrier into a different well, with a new barrier~$E'$.
Thus yielding can be identified with barrier crossing.
This description is similar to that of activated
processes in thermodynamic systems;
however, the random fluctuations here are
{\em not necessarily} due to the true thermodynamic temperature.
They may arise rather from a form of homogeneous mechanical noise
generated by non--linear couplings between the elements;
see \cite{sgr_pre} for a fuller discussion on this point.
To avoid possible confusion,
this effective temperature is denoted by the symbol~$x$
rather than~$T$.

Although the energy barrier of an element is initially~$E$,
as it becomes strained it will gain an elastic energy of
$\frac{1}{2}kl^{2}$ and thus will have a smaller effective energy barrier
$\eff=E-\frac{1}{2}kl^{2}$.
Thus the yield rate will take the form

\begin{equation}
\Gamma(E,l)=
\Gamma_{0}\exp\left[-\:\frac{E-\frac{1}{2}kl^{2}}{x}\right]\quad,
\label{e:gamma}
\end{equation}

\noindent{}where the attempt rate $\Gamma_{0}$ sets the
time scale of the yielding.
The effective temperature $x$ is constant in the SGR model,
and essentially acts as a parameter of the model.
However, since $x$ may be generated in part
by internal couplings between the elements,
it should be allowed to vary with the state of the system,
{\em i.e.} $x=x(\{P(E,l,t)\})$.

In this paper, we shall not consider the most general
possible form for~$x$, but shall instead
focus on a more restricted class for which $x=x(\sigma,l)$.
This corresponds to the realisation that,
as an element is strained, it may become more or
less susceptible to the noise, and hence its
`temperature' $x$ may change.
Allowing $x$ to also depend on $\sigma$
reflects that this change in susceptibility to noise
may in part be a {\em global} phenomenon,
{\em i.e.} the $x$ of a given element may depend on
the state of all of the elements around it.
Clearly, deriving the actual $x(\sigma,l)$
for any given material would be a highly complicated task,
of comparable difficulty to the original coarse--graining
procedure described above.

For further simplicity,
we shall not consider $x=x(\sigma,l)$
but shall instead focus on two limiting cases~:
$x=x(\sigma)$ is described in Sec.~\ref{s:global},
and
$x=x(l)$ is assumed in Secs.~\ref{s:local} and~\ref{s:monodisperse}.
The relevant results for the SGR model, which corresponds
to the case of $x=$(constant),
are also summarised in Sec.~\ref{s:sgr}.

\subsection{Natural units and the $\rho(E)$ distribution}
\label{s:prior}

We have yet to specify $\rho(E)$, the distribution
of energy barriers $E$ for elements that have just yielded.
In the SGR model, $\rho(E)$ is assumed to have an exponential
tail, $\rho(E)\sim{\rm e}^{-E/E_{0}}$ as $E\to\infty$,
which gives rise to a finite
yield stress and diverging viscosity
for some values of $x$, but not others~\cite{sgr_pre}.
Although it is possible to justify the occurrence of
an exponential tail in some contexts~\cite{jpb_mm},
we prefer instead to treat this choice of $\rho(E)$,
combined with the Arrhenius form of $\Gamma(E,l)$~(\ref{e:gamma}),
as a recipe for generating a yield stress within
this simple picture of activated yielding.
This may seem {\em ad hoc}, but it should be realised
that jamming is in reality a many body effect,
involving collective behaviour between
a large number of degrees of freedom.
It should therefore come as no surprise that
assumptions are required to describe jamming
within this single--particle picture.

For much of this paper, we shall assume that
$\rho(E)$ has an exponential tail, just as in the SGR model.
In fact, for the numerics the definite form
$\rho(E)=\frac{1}{E_{0}}{\rm e}^{-E/E_{0}}$ has been used,
although it is not expected that the precise shape of
$\rho(E)$ for small $E$ will
alter the long--time or steady state behaviour of the system.
This is because small values of $E$ correspond to elements with short
expected lifetimes.
Furthermore, it turns out that the oscillatory motion
described in Sec.~\ref{s:local} does {\em not} depend
on the choice of $\rho(E)$, and in Sec.~\ref{s:monodisperse}
the simpler form $\rho(E)=\delta(E-E_{1})$ is used,
with qualitatively similar results.

There remain three constants in the problem.
These are the elastic constant~$k$,
the attempt rate~$\Gamma_{0}$ in~(\ref{e:gamma}),
and the constant $E_{0}$ in the definition of $\rho(E)$ given above.
However, these can all be scaled out of the model.
For instance, $k$~sets the scale of the stress, and therefore can
be removed by rescaling $\sigma$ to $\sigma/k$.
Similarly, $\Gamma_{0}$ gives the only intrinsic time scale in the system,
and can be scaled out through $t$ and~$\gd$.
Finally, $E_{0}$ sets the scale of the energy barriers,
and can be scaled out through~$x(\sigma,l)$.

In what follows, we have adopted {\em natural} units
in which $k=\Gamma_{0}=E_{0}=1$.
This means that the actual values for $t$, $\gamma$ and $\sigma$
given below should not be compared to experimental
values without first rescaling with the appropriate
$k$, $E_{0}$ and $\Gamma_{0}$ for the material in question.
For example, the typical yield strain according to
the $\Gamma(E,l)$ given in (\ref{e:gamma}) is $\sqrt{2E_{0}/k}$,
which is when the effective energy barrier vanishes.
In natural units, this is of $O(1)$; however,
for soft materials it will typically be of only a few percent,
and will be even smaller for hard materials.
Thus the scale of the local strains~$l$,
and therefore the scale of the global strain~$\gamma(t)$,
will generally be significantly smaller in real materials
than with natural units.

\section{Summary of the standard SGR model}
\label{s:sgr}

As discussed in the previous section,
the standard SGR model is realised when
(using natural units) the yield rate
$\Gamma(E,l)$ is chosen to take an Arrhenius form with
an energy barrier $E-\frac{1}{2}l^{2}$ and a constant
effective temperature~$x$,
and $\rho(E)$ has an exponential tail $\rho(E)\sim{\rm e}^{-E}$.
Many results for this case are already
known~\cite{sgr_prl,sgr_pre,sgr_aging}.
The purpose of this section is to briefly describe those
results that will be referred to in later sections.

\begin{figure}
\centerline{\psfig{file=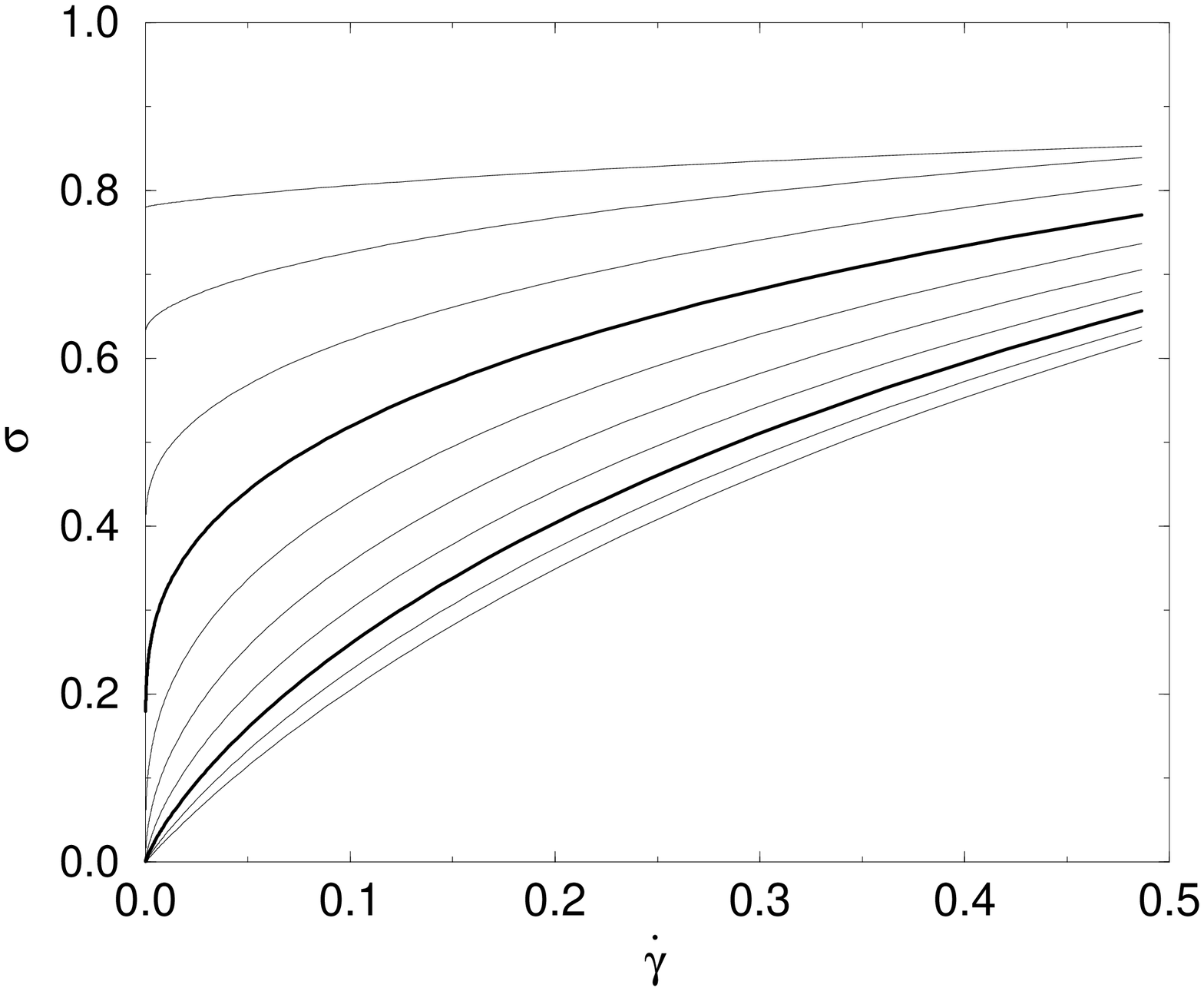,width=9cm,angle=270}}
\caption{Flow curves for a system with a constant~$x$,
{\em i.e.} the SGR model.
From top to bottom, each line corresponds to a value of $x$
increasing from 0.25 to 2.5 in steps of 0.25.
The lines $x=1$ and $x=2$ have been highlighted.
A finite yield stress $\yield\equiv\lim_{\gd\to0}(\sigma)$
exists only for $x<1$.
The prior distribution is $\rho(E)={\rm e}^{-E}$.}
\label{f:sgr_flow_curves}
\end{figure}

Let us assume that the system is driven by a given strain
$\gamma(t)$, and that $\gamma(t)\sim\gd t$ as $t\rightarrow\infty$,
where $\gd\neq0$.
Without loss of generality we take $\gd>0$ hereafter.
Under these conditions, the system will reach
the steady state solution $P_{\infty}(E,l)$ already given
in~(\ref{e:st3}).
The asymptotic stress $\sigma$ is found by averaging
the local strain $l$ over $P_{\infty}(E,l)$,
{\em i.e.} $\sigma=\langle l\rangle$,
which defines the flow curve $\sigma(\gd)$.
Example flow curves for different values of $x$
are shown in Fig.~\ref{f:sgr_flow_curves}.
In all cases the gradient
${\rm d}\sigma/{\rm d}\gd$
decreases with
increasing~$\gd$, indicating that the
apparent viscosity $\eta\equiv\sigma/\gd$ decreases with $\gd$
and that the system is everywhere shear thinning.

It should also be noted that there is a
a one--to--one correspondence
between $\gd$ and~$\sigma$.
This means that every point on the curve
can be reached in one of two ways~: either by applying
a known strain $\gamma(t)\sim\gd t$, as described above,
or by imposing a constant stress $\sigma$.
The uniqueness of the steady state solution ensures
that the same $P_{\infty}(E,l)$ will be reached in both cases.
Of course, this assumes that the steady state is reached at all,
for which it must both exist and be stable.
For the standard SGR model, the steady state solution is
stable as long as it exists~\cite{sgr_pre,sgr_aging},
but this does not hold for all $x(\sigma,l)$
under an imposed stress,
as will be discussed in later sections.

The behaviour of $\sigma$ as $\gd\rightarrow0^{+}$ depends upon
the choice of~$x$~\cite{sgr_pre}.
For instance, for $x>2$ the stress scales as
$\sigma\sim\gd$ and the system is Newtonian,
whereas there is a power law fluid regime
$\sigma\sim\gd^{x-1}$ for $1<x<2$.
However, the most important regime for our purposes
is $x<1$, for which $\sigma$ approaches a finite value.
The yield stress $\yield$, defined by

\begin{equation}
\yield
=
\lim_{\gd\rightarrow0^{+}}
\sigma(\gd)\quad,
\end{equation}

\noindent{}smoothly tends to zero as $x\rightarrow1^{-}$
and remains zero for all $x\geq1$.
If a system with $x<1$ has a stress
$\sigma<\yield$ applied to it,
then it will never reach a steady state with a finite~$\gd$,
simply because $P_{\infty}(E,l)$ for these parameters
does not exist.
Instead the strain will logarithmically creep according
to $\gamma(t)\sim\ln t$~\cite{sgr_aging}.
There may be a crossover to a different
behaviour at late times if $\sigma$ is close to $\yield$,
but it must always be true that $\gd\to0^{+}$.

The logarithmic creep in $\gamma(t)$ under an imposed
stress $\sigma<\yield$ is an important result.
If it is realised in an experimental situation,
then when the strain rate $\gd\sim1/t$
drops below the resolution of the
apparatus, which it must do at long times, it might be erroneously
deduced that the system has stopped flowing altogether,
{\em i.e.} $\gd\approx0$.
This is in accord with our intuitive notions about jamming~:
the sample initially flows when pushed, but later stops
flowing, or `jams.'
One might argue that, technically, a jammed system should
have $\gd=0$ exactly, but it is not possible for any
model in this class to sustain a finite stress without
a finite~$\gd$
(unless the sample is allowed to age for an arbitrarily long
period of time before being sheared; see~\cite{sgr_aging}).
This is because each individual element
has a finite relaxation time until it yields and releases
its stress, which can only be balanced by an increasing strain.

With this insight, we now {\em define} a `jammed' state for
this class of systems to be a configuration
which has a finite limiting stress $\yield$
when it is driven by an arbitrarily small strain rate $\gd=0^{+}$.
For the SGR model, this corresponds to $x<1$;
however, for $x=x(\sigma)$ it may also depend on the
history of the sample, as will now be discussed.

\section{Jamming model A: \lowercase{$x=x(\sigma)$}}
\label{s:global}

It is useful to recall the physical picture that
underlies an $x=x(\sigma)$.
As the system becomes sheared, either by an imposed
strain $\gamma$ or an imposed stress~$\sigma$,
it will become distorted, and this may affect
its susceptibility to noise.
The precise manner in which this happens will
depend on the microscopic composition of the material
in question.
For the systems we are interested in,
{\em i.e.} those that can shear--thicken or jam,
the distorted configuration is {\em less} susceptible to noise
that the undistorted one.
One way to incorporate this effect is to
regard the system as becoming `colder' as its shear stress increases,
which corresponds to a decreasing~$x(\sigma)$.
Thus setting $x=x(\sigma)$ provides a mechanism for allowing the effective
temperature $x$ to evolve with the global state of the system.

In this section we shall consider $x(\sigma)$ that
takes values greater than 1 for some ranges of~$\sigma$,
and less than 1 for others.
According to the results of the previous section,
this should allow the system to change from a jammed
to a flowing state in response to the driving,
or more precisely, in response to changes in~$\sigma(t)$.
Thus we may be able to observe a shear--induced jamming transition
(something that is not possible in the SGR model,
for which the existence of a yield stress depends
purely on the choice of the external parameter~$x$).
For clarity, we shall restrict our attention to the simplest
choice of $x(\sigma)$ that has the potential to
shear--thicken and jam,
namely $x(\sigma)>1$ for small $\sigma$ and $x(\sigma)<1$ for
large~$\sigma$,
with a monotonic (possibly discontinuous)
behaviour for intermediate~$\sigma$.

\subsection{Steady state behaviour}
\label{s:global_ss}

\begin{figure}
\centerline{\psfig{file=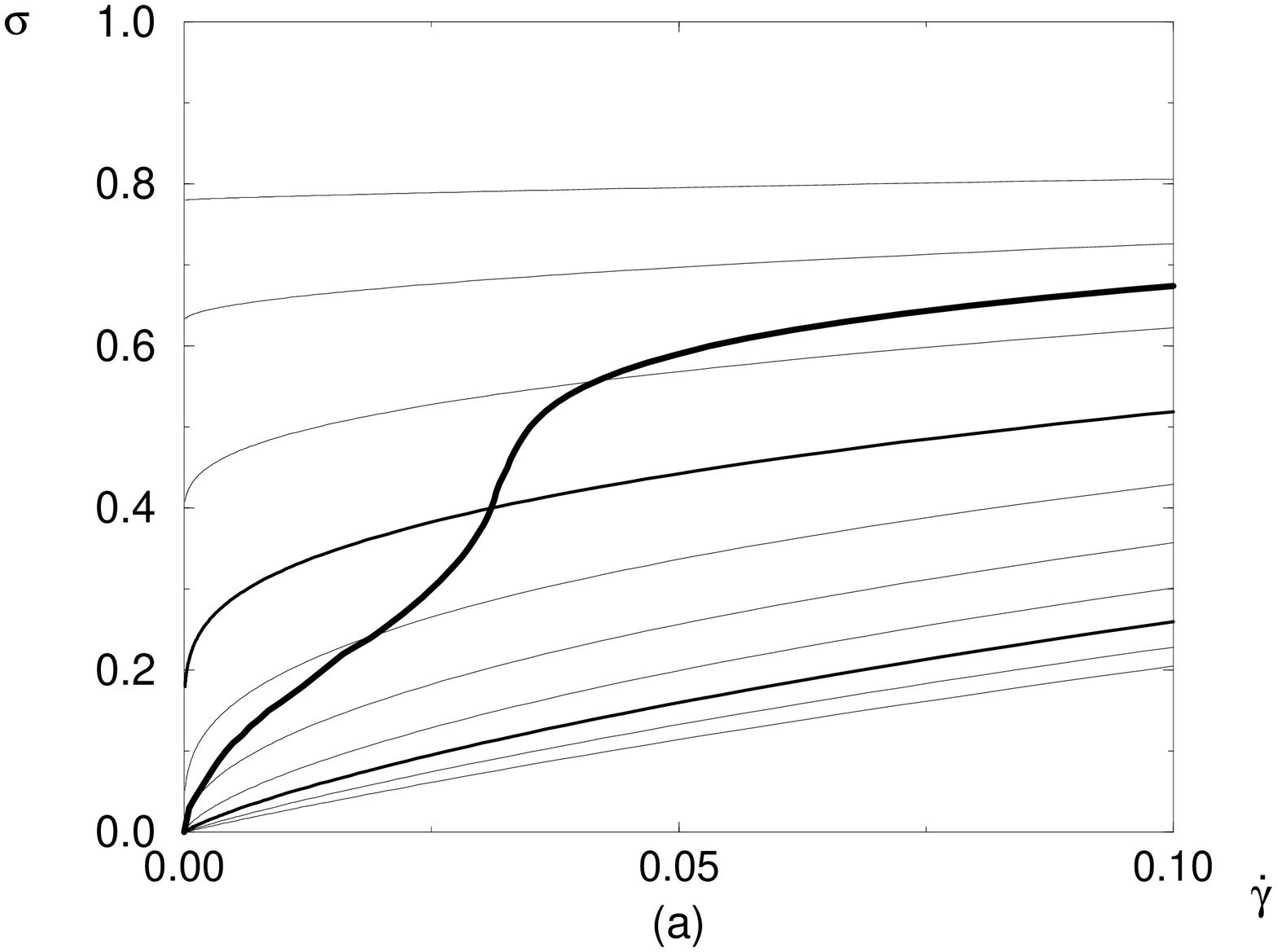,width=9cm,angle=270}}
\centerline{\psfig{file=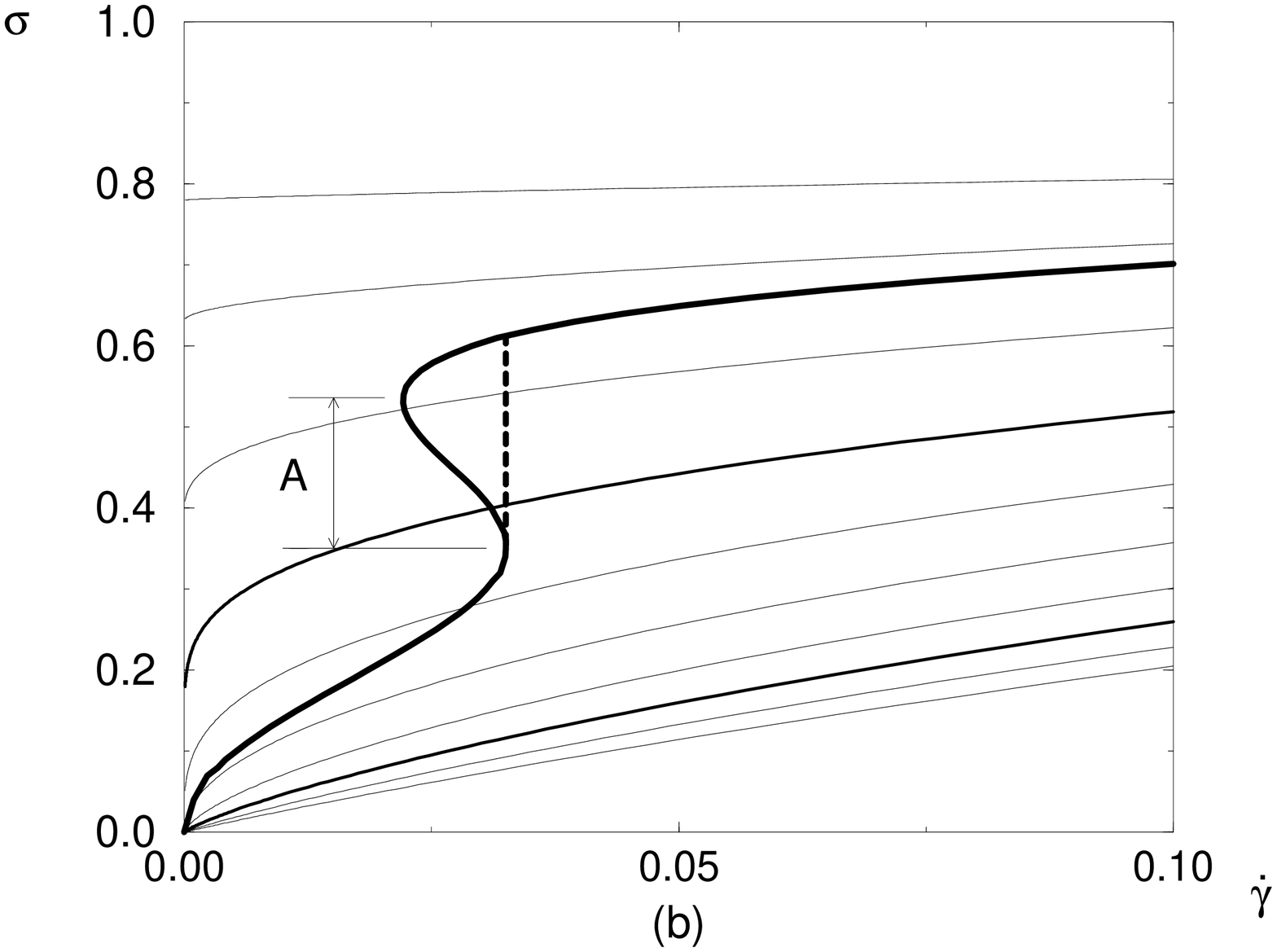,width=9cm,angle=270}}
\centerline{\psfig{file=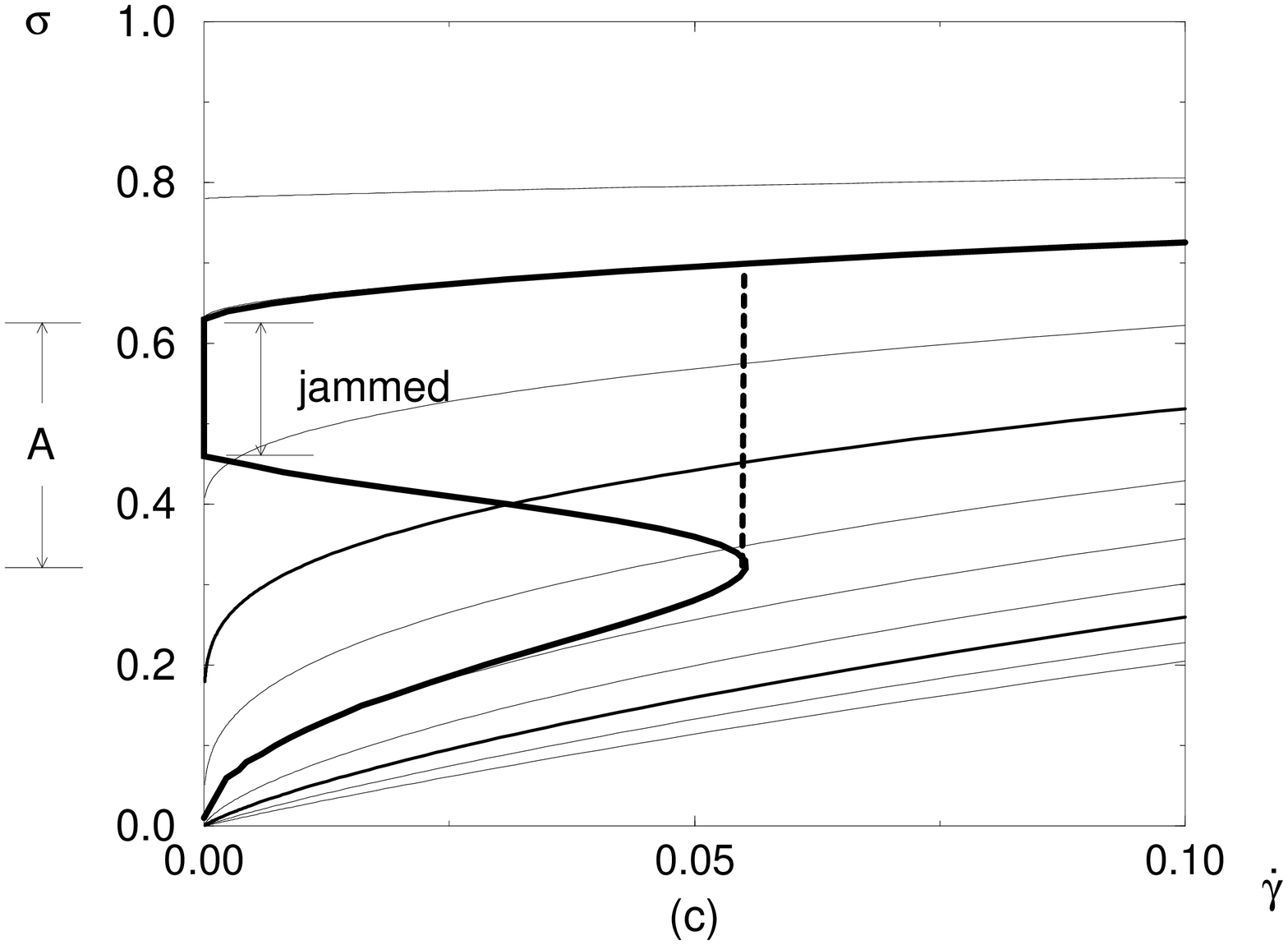,width=9cm,angle=270}}
\centerline{\psfig{file=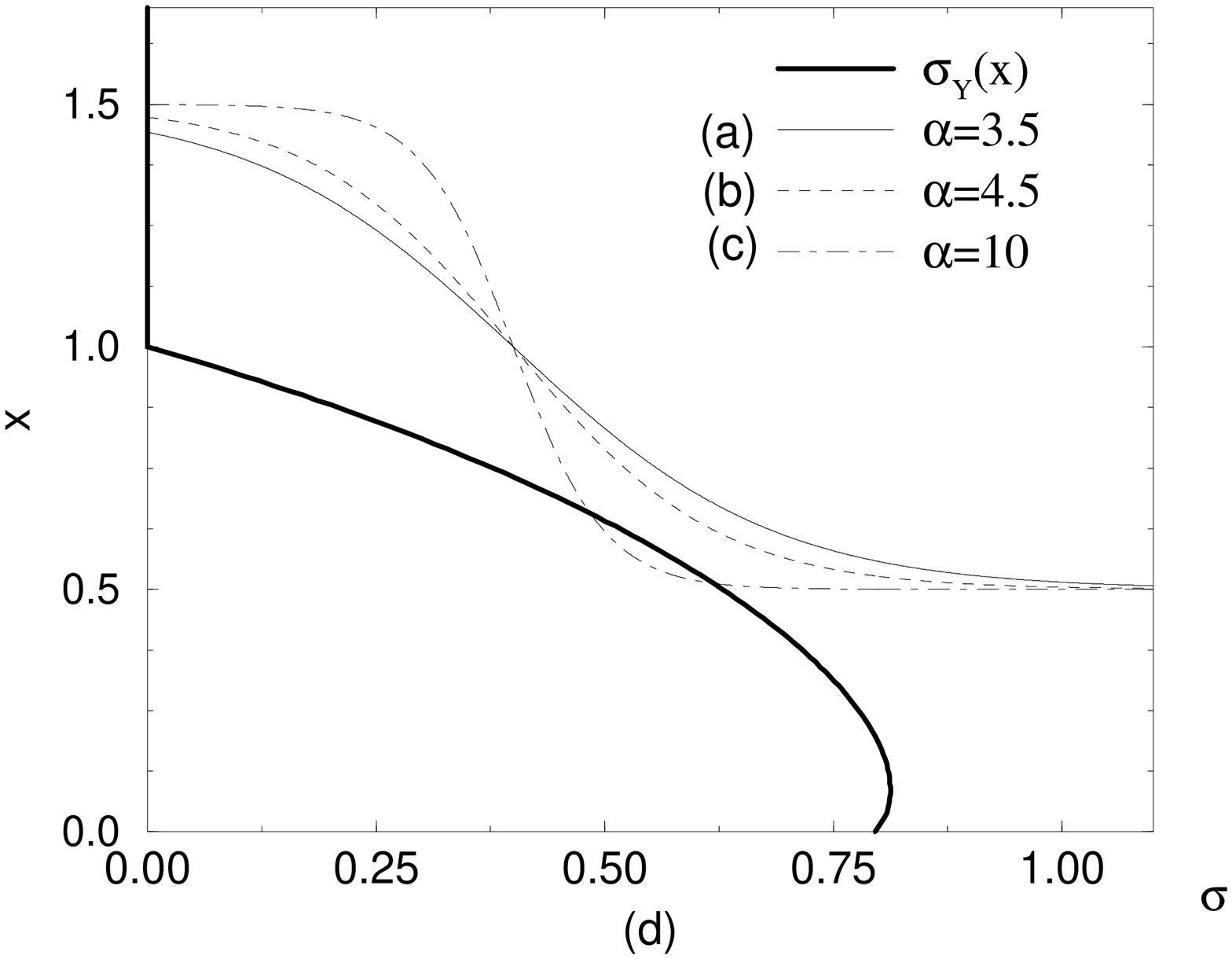,width=9cm,angle=270}}
\caption{Flow curves for
$x(\sigma)=1-0.5\tanh[\alpha(\sigma-0.4)]$ (thick lines),
where $\alpha=3.5$~(a), 4.5~(b) and 10~(c).
For comparison, the thin lines are the constant $x$ curves from
Fig.~\ref{f:sgr_flow_curves}.
In (b) and (c), the vertical dashed line represents
the discontinuous jump in stress for a slowly increasing~$\gd$,
and the region marked `A' denotes the range of $\sigma$
that is unstable under an imposed~$\gd$, but is stable
under an imposed~$\sigma$.
In (c), the range of $\sigma$ for which the system is `jammed,'
{\em i.e.} $\sigma(0^{+})>0$, has also been marked.
The $x(\sigma)$ for different $\alpha$ are plotted in~(d),
where it can clearly be seen that only for the
$\alpha=10$ case does
$x(\sigma)$ drop below the yield stress curve.
}
\label{f:jamming_curves}
\end{figure}

Although allowing $x$ to vary in time
according to $x=x[\sigma(t)]$
can complicate the transient behaviour of the system,
as described below,
the steady state is easy to analyse.
This is because the very definition of a steady state means that
$\sigma$ asymptotically approaches a constant value,
and thus $x$ also becomes constant.
There is therefore a straightforward procedure
for generating the $x=x(\sigma)$ flow curves from those
of constant~$x$~:
for any given value of~$\sigma$,
one simply finds the SGR flow curve for the corresponding
value of $x(\sigma)$ and reads off the required value of~$\gd$.
This amounts to interpolating between the various SGR flow curves.
Some examples are given in Fig.~\ref{f:jamming_curves}
for various $x(\sigma)$ that change from 1.5 for small $\sigma$
to 0.5 for large~$\sigma$.

If $x(\sigma)$ takes values greater than 1 for small $\sigma$,
but smaller than 1 for larger $\sigma$, then it may jam
under an applied stress.
However, for this situation to be realised,
the applied stress must be simultaneously
large enough that $x(\sigma)<1$,
and also small enough that $\sigma<\yield$\,.
Since the yield stress $\yield$ also depends on~$x$,
the precise criterion for the upper yield stress to be attainable
is that there is a range of $\sigma$
for which

\begin{equation}
\sigma<\yield[x(\sigma)]\quad.
\label{e:criterion}
\end{equation}

\begin{figure}
\centerline{\psfig{file=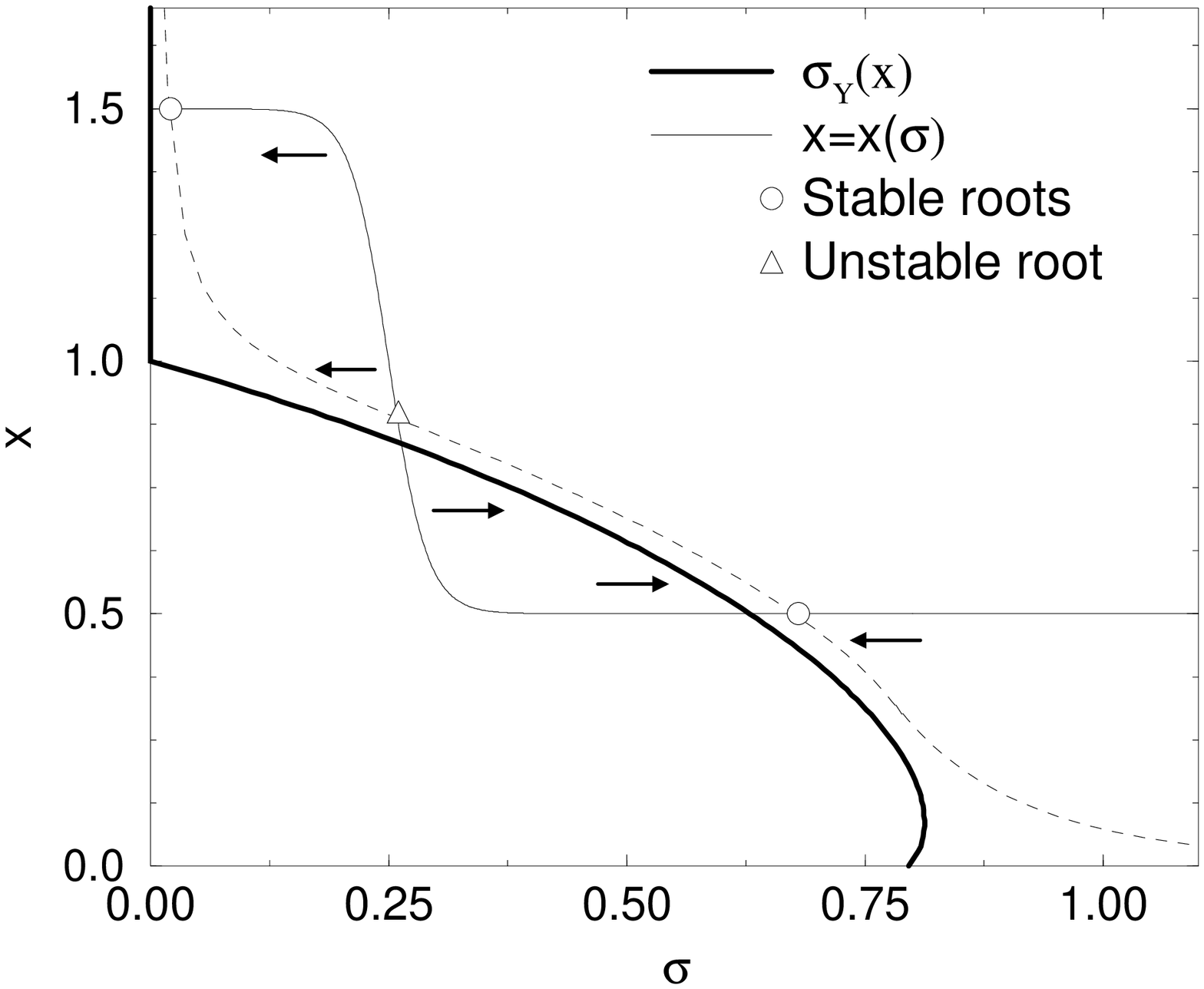,width=10cm,angle=270}}
\caption{Plot of the yield stress $\yield(x)$ from the SGR model,
overlayed with a particular choice of $x(\sigma)$
that changes value from 1.5 to 0.5 with increasing stress
(the actual functional form is
$x(\sigma)=1-\frac{1}{2}\tanh[25(\sigma-0.25)]$).
The dashed line is a schematic representation of
the asymptotic stress $\sigma(t\rightarrow\infty)$
for a small but finite~$\gd$,
which is what is actually attainable for these models.
The arrows represent the direction in which the stress will vary
for a constant~$x$, and explain the given stability of the roots.
}
\label{f:roots}
\end{figure}

\noindent{}This corresponds to the region in which
the $x(\sigma)$ curve falls below the $\yield(x)$ curve
when they are both plotted on the same axes,
as in Fig.~\ref{f:roots}.
For example,
of the flow curves already presented in Fig.~\ref{f:jamming_curves},
only the third example obeys the criterion~(\ref{e:criterion})
for a finite range of $\sigma$.
Imposing a stress $\sigma$ from within this range will result
in a system that is jammed in the sense that the strain
logarithmically creeps, $\gamma(t)\sim\ln t$
and $\gd(t\to\infty)=0$.

Two additional complications arise when the system is driven
by an imposed strain rate $\gd$ rather than an imposed stress.
Firstly,
if $\gd$ is increased at a sufficiently slow rate that the system
is always arbitrarily close to its steady state,
the stress may undergo a discontinuous jump
from one branch of the flow curve to another,
as marked in Figs.~\ref{f:jamming_curves}(b) and \ref{f:jamming_curves}(c).
If $\gd$ is decreased in a similar fashion,
then $\sigma(t)$ will follow the upper branch
until it again jumps discontinuously at a {\em different}
value of~$\gd$.
Thus the system exhibits {\em hysteresis}.
Hysteresis due to non--monotonic flow curves has also been observed 
experimentally~\cite{volkova,bonn,jacques,laun}.
The complementary form of non--monotonicity,
which would allow a discontinuous jump in $\gd$ for
a small change in~$\sigma$, cannot be realised in this class
of models.
This is because there is only one value of $x(\sigma)$,
and hence one steady state solution,
for any given value of~$\sigma$.

The second complication concerns the stability of the jammed state.
In Fig.~\ref{f:roots}, the yield stress $\yield(x)$
and an $x(\sigma)$ that obeys (\ref{e:criterion}) for a range
of $\sigma$ have been plotted on the same axes.
Also plotted is the line of $\sigma$ that can be
reached for a small but finite~$\gd>0$,
which converges to $\yield(x)$ as $\gd\to0$.
These are the $\sigma$ that can be realised in practice,
since the steady state solution (\ref{e:st3})
is not defined for $\gd\equiv0$.
For this example there are 3 roots~:
a `flowing' root with $\sigma=0^{+}$, and two `jammed'
roots with finite $\yield>0$.
The arrows in this diagram
point in the direction in which the stress will be
varying for a given point on the line $x(\sigma)$.
This direction is based on the assumption that, for time scales over
which $x$ can be treated as a constant,
which will certainly include the asymptotic regime,
the system will behave like the SGR model and
$\sigma$ will evolve towards $\yield$.
It is clear from this that all points tend to flow away
from the middle root, which is therefore unstable.
If the system is initially placed near to this root,
it will undergo a transient and converge to either the higher root,
or to the `flowing root' with $\sigma=0^{+}$.


Fig.~\ref{f:roots} can also be used to predict the
stability of the steady state for finite~$\gd$.
As $\gd$ increases,
the asymptotic stress $\sigma(\gd)$ for the SGR model
will move away from the yield stress curve,
but will always remain a monotonic decreasing
function of~$\gd$~\cite{sgr_pre}.
Thus a smoothly decreasing~$x(\sigma)$ will only intersect with
it 1 or 3 times;
when there are 3 roots, the middle one is always unstable,
for precisely the same reasons as given in the previous
paragraph.
The range of $\sigma$ which are unstable and
cannot be realised under
an imposed $\gd$ have been labelled `A' in
Figs.~\ref{f:jamming_curves}(b) and~(c).
Note that the unstable roots correspond to
regions of the flow curve with a negative gradient,
as in experiments and from other theoretical
considerations~\cite{olmsted,lu,volkova,bonn,jacques,hu}.
There are no stability issues for an imposed~$\sigma$,
for which there is always a single, stable root.

\subsection{Transient behaviour under an imposed \lowercase{$\gd$}}

There are in principle no difficulties in extending the results
described above to a system that is driven by
a time--dependent imposed stress $\sigma(t)$.
As long as $\sigma(t)$ tends to a constant value $\sigma_{0}$
at long times, then the same steady state behaviour
as previously discussed will apply, with $\sigma$ replaced
by~$\sigma_{0}$\,.
The transient behaviour does not affect the final state.
However, this is not the case when the system is instead
driven by an imposed strain rate~$\gd(t)$.
In this case, $\sigma(t)$ can vary in a manner that is difficult
to predict in advance, so that the final stress reached,
and hence whether or not the system is jammed,
will in general depend on the entire history of the system.

\begin{figure}
\centerline{\psfig{file=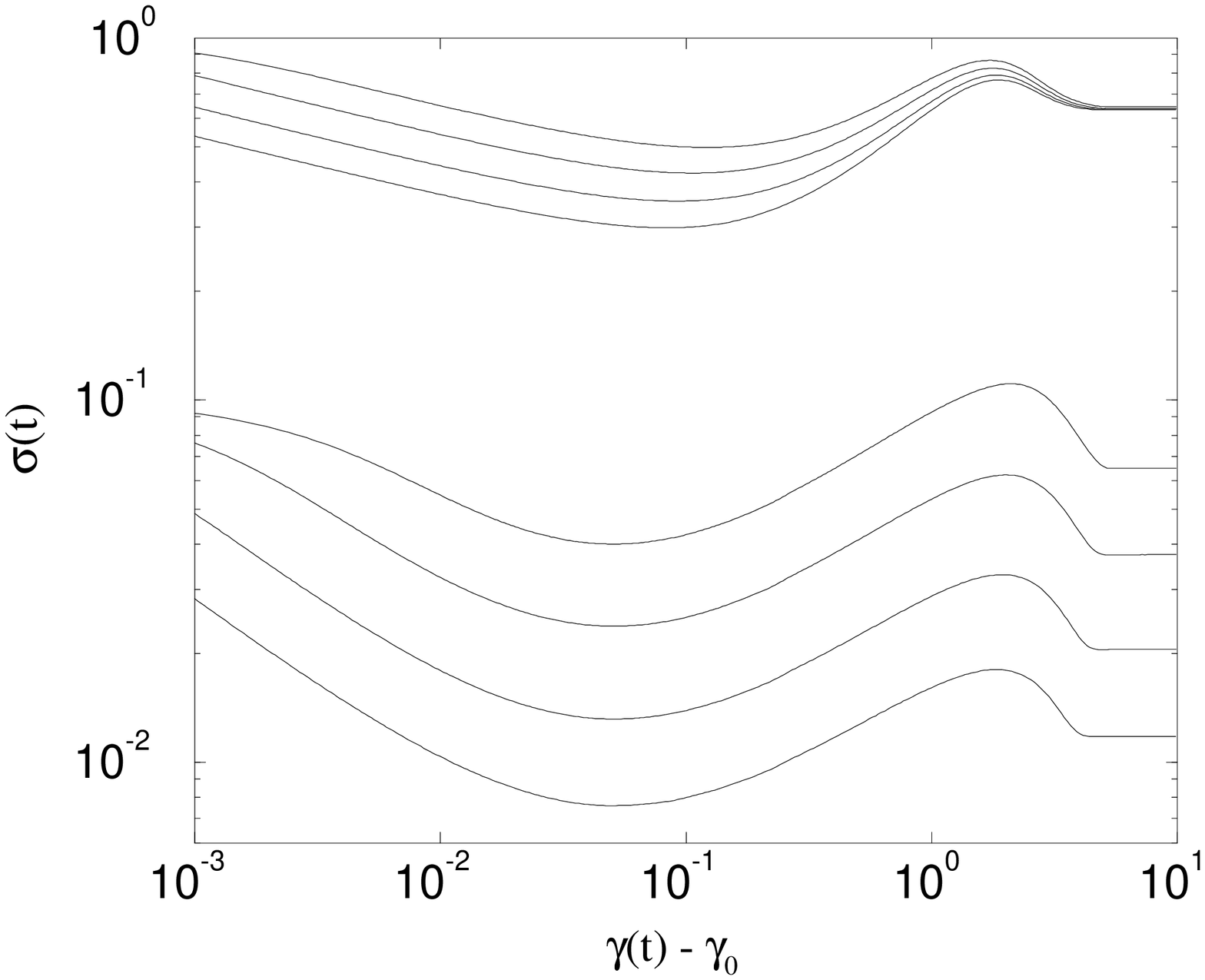,width=10cm,angle=270}}
\caption{Plot of stress versus strain for $\gamma(t)=\gamma_{0}+\gd t$
and the same $x(\sigma)$ as in Fig.~\ref{f:roots}.
The upper set of lines correspond to $\gamma_{0}=1$ and the
lower set to $\gamma_{0}=0.1$.
Within each set, the lines refer to (from top to bottom)
$\gd=3\times10^{-3}$, $10^{-3}$, $3\times10^{-4}$ and $10^{-4}$.
As $\gd\to0^{+}$,
the upper curves are seen to be converging to
$\sigma\to\yield\approx0.65$,
whereas the lower curves are approaching a zero--stress
state according to $\sigma\approx k\gd^{0.5}$,
where $k$ is an arbitrary constant.
In all cases the system was first allowed to reach equilibrium under
zero shear before the step shear $\gamma_{0}$ was applied.
}
\label{f:existence_of_yield}
\end{figure}

A concrete example of the history dependence of a yield stress
is given in Fig.~\ref{f:existence_of_yield}.
Here, the system is first subjected to a step shear
of magnitude $\gamma_{0}$, and is then continuously sheared
at a rate~$\gd$, {\em i.e.} $\gamma(t)=\gamma_{0}+\gd t$.
As $\gd$ tends to zero, the stress approaches an asymptotic value
$\yield$; however, whether or not $\yield$ is finite
depends on~$\gamma_{0}$.
For $\gamma_{0}=1$ the stress is seen to be converging to
a finite value $\sigma\approx0.65$, but for $\gamma_{0}=0.1$
it rather tends to a `flowing' state with $\sigma=0^{+}$
(more precisely, $\sigma\propto\gd^{0.5}$ as $\gd\to0^{+}$).
The system can be said to exhibit strong long--term memory,
where the use of the word `strong' means that the memory
of an earlier perturbation of finite duration
does not decay to zero with time~\cite{sgr_aging,glassrev,weak1,weak2}.

\begin{figure}
\centerline{\psfig{file=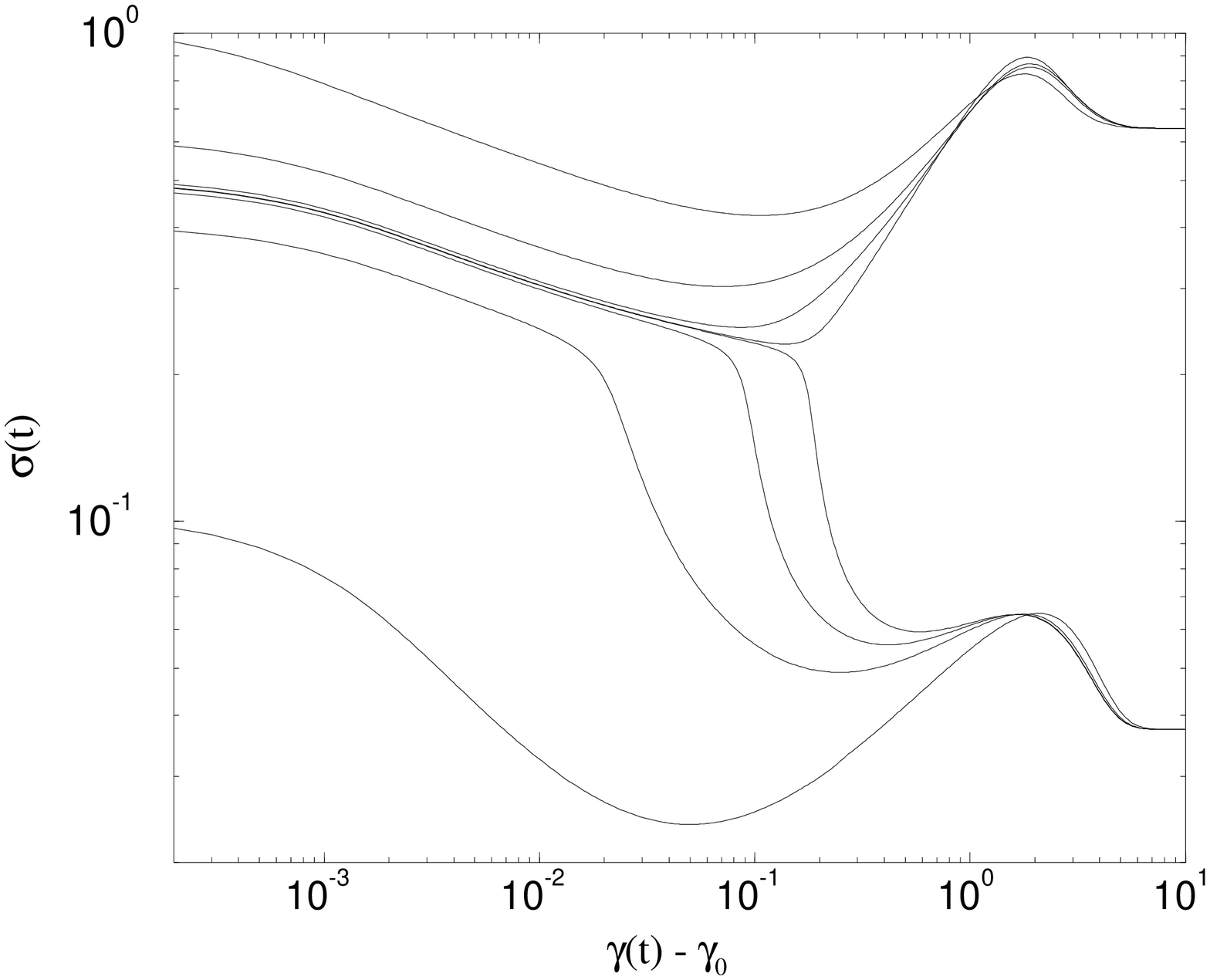,width=10cm,angle=270}}
\caption{The variation in stress for the system of
Fig.\ref{f:existence_of_yield} driven by the
imposed strain $\gamma(t)=\gamma_{0}+\gd t$ with $\gd=10^{-3}$,
demonstrating the inability to reach the root
at $\sigma\approx0.3$.
From bottom to top on the left hand side,
the `initial condition' $\gamma_{0}$ takes the values
$\gamma_{0}=0.1$, 0.4, 0.48, 0.49, 0.491, 0.5, 0.6 and~1.
}
\label{f:missed_root}
\end{figure}

The choice of $x(\sigma)$ used in the previous example is in
fact the same as that plotted in the stability diagram,
Fig.~\ref{f:roots}.
This allows for a striking illustration of the instability
of the middle root that lies near the point $\sigma\approx0.3$
on the stability diagram.
Plotted in Fig.~\ref{f:missed_root} are the $\sigma(t)$
curves for a fixed $\gd=10^{-3}$ and a range of values
$0.1\leq\gamma_{0}\leq1$.
It is clear that the stress will always diverge away from
the unstable root at late times, no matter how
finely $\gamma_{0}$ is tuned.
Thus only the roots at $\sigma=0^{+}$ and $\sigma\approx0.65$
are stable, as previously claimed.

There are many other complications that can arise due transient
behaviour in a strain--controlled system.
For instance, all of the $\sigma(t)$ curves
plotted in Fig.~\ref{f:existence_of_yield}
reach their global minimum $\sigma_{\rm min}$
at a time $1\ll t\ll1/\gd$.
It can be shown that $\sigma_{\rm min}$ becomes
arbitrarily small as $\gd\to0^{+}$.
Since this corresponds to a state with a high effective temperature
$x(\sigma_{\rm min}=0^{+})>1$
in which the stress is rapidly dissipated,
then if $\gd$ is sufficiently small,
$\sigma(t)$ will remain low and the system will crossover to
the flowing root, irrespective of~$\gamma_{0}$.
However, this effect can be reversed if the initial step shear
is replaced by a smoothly varying $\gd(t)$,
such as one that exponentially decays towards its final value of~$\gd$,
which is more like what could be attained in
an actual rheometer.
We will not pursue these complications any further here.

In summary, the SGR model generalised to allow
the effective temperature $x$ to vary with the global stress~$\sigma$
can exhibit, for suitable choices of $x(\sigma)$~:
{\em hysteresis} in $\sigma(t)$ for slowly varying $\gd$,
as seen from the flow curves in Fig.~\ref{f:jamming_curves};
shear induced {\em jamming}, as in Fig.~\ref{f:existence_of_yield};
and
{\em strong history--dependence} of the existence
of a yield stress.

\section{Jamming model B: \lowercase{$x=x(l)$}}
\label{s:local}

The second limiting form for $x=x(\sigma,l)$ to be considered
is one that depends only on the local strain, $x=x(l)$.
This marks a more drastic departure from the SGR model
than the $x(\sigma)$ considered in the previous section,
since now every element will generally have a different
effective temperature~$x$.
The steady state will therefore be different to that of
the SGR model,
and it will not now be possible to map the
flow curves for $x(l)$ across from those for constant~$x$.
One could also envisage regions of the parameter
space for which the steady state cannot even be reached.
Indeed, this is precisely what we find~: for some finite region
of parameter space,
the system reaches an oscillatory regime under an imposed
stress, but not under an imposed strain.
This is the central finding of this section.

The physical justification in choosing $x=x(l)$ is similar to
that already discussed for $x=x(\sigma)$, {\em i.e.} it is assumed
that the material can become more or less susceptible to
noise in its strained state.
The main difference is that this is now assumed to be a
local effect, that can be described purely on the level of
individual elements.
Just as in the previous section,
we shall focus our attention on $x(l)$ that decrease with~$l$,
since it is these that have the potential to
exhibit shear thickening.

\subsection{Monotonicity of the flow curves}

\begin{figure}
\centerline{\psfig{file=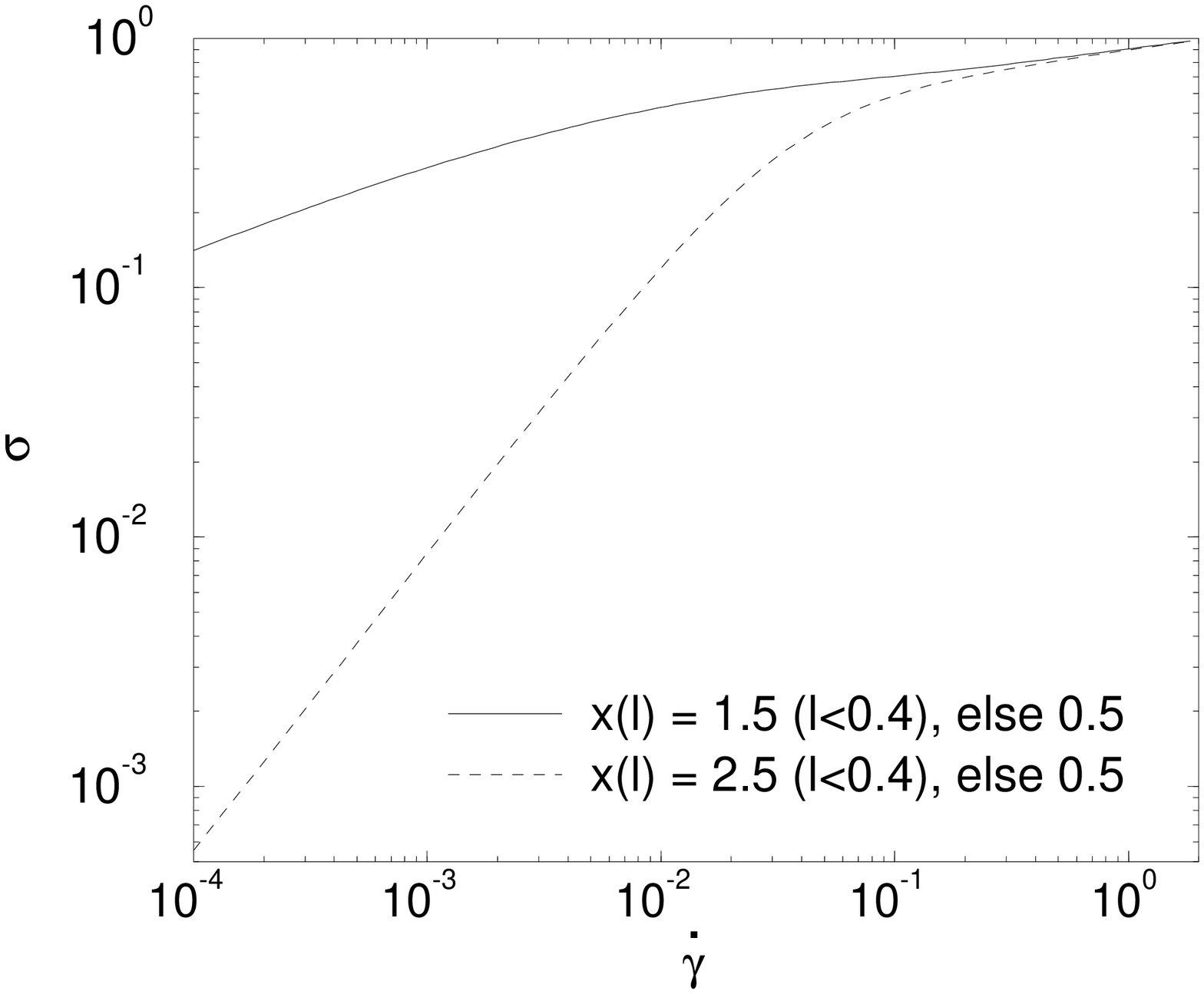,width=9cm,angle=270}}
\caption{Examples of typical flow curves when $x$ depends on
the local strain~$l$.
Here, $x(l)=0.5$ for $l>0.4$, but takes the higher value
of 1.5 (solid line) or 2.5 (dashed line) for $l<0.4$.
These lines were generated from the steady state solution
of the master equation.
}
\label{f:local_flow_eg}
\end{figure}

Given $x(l)$, the steady state stress $\sigma$ for a given $\gd$ can be
found by calculating the mean strain $\langle l\rangle$
for the steady state solution~(\ref{e:st3}).
Some example flow curves are plotted in Fig.~\ref{f:local_flow_eg}.
They are clearly monotonic~: their gradient is
everywhere positive, and there is a 
one--to--one relationship between $\sigma$ and $\gd$.
In these figures, the $x(l)$ were chosen to vary in a stepwise manner,
taking a value $x>1$ for $l<0.4$
and a lower value $x<1$ for $l>0.4$.
This is typical of the $x(l)$ employed in this section.
However, the monotonicity result is entirely general
and applies for any $x=x(l)$,
as demonstrated in Appendix~\ref{s:monotonicity}.

The physical reason for the monotonicity of the flow curves
is that the elements are uncoupled when $x=x(l)$
and the system is driven by an imposed strain~$\gamma(t)$.
That is, the expected strain reached before a given
element yields does not depend on the state of any other elements.
This is {\em not} true for the general case $x=x(\sigma,l)$,
when the value of $x$ for an element depends on the global
stress $\sigma$, and thus on the state of the rest of the system.
As long as independence holds,
the average strain reached
before each individual element yields,
and hence the global stress~$\sigma$,
can only increase with increasing~$\gd$.
Thus the flow curves must be monotonic.

The monotonicity of the flow curves is an important finding.
As mentioned in the introduction, rheological instabilities
often arise when the system has been driven
to a point on its flow curve which has a negative gradient.
Fluctuations can then cause spatial inhomogeneities to arise,
such as shear bands with either the same stress and different strain rates,
or equal strain rates but differing stresses~\cite{olmsted,lu}.
Alternatively, temporal oscillations may be observed~\cite{jacques,hu}.
Since the flow curves for $x=x(l)$ have no regions of negative
slope, one would not expect any instabilities to appear in this model.
Nonetheless temporal oscillations in $\gd$ do occur
for a finite range of imposed~$\sigma$,
as we now describe.

\subsection{Oscillatory behaviour under an imposed $\sigma$}
\label{s:poly_osc}

The monotonicity result described above was attributed to
the independence of the elements, which is
only true for a strain controlled system.
By contrast,
the elements become coupled under an imposed $\sigma$
since, for example, a single element yielding causes the
mean strain to decrease slightly, which must be immediately
countered by an increase in $\gd$,
{\em which affects every element in the system}.
Thus collective behaviour can now occur.
In fact, this collective behaviour cannot alter any system
that has already reached steady flow,
which we know is identical for strain and stress controlled systems.
Thus as long as the stress controlled system reaches steady flow,
it will fall onto the same monotonic flow curve as before.
However, the couplings between the elements can drastically
alter the transient behaviour, to such an extent that
steady flow may never be reached.

\begin{figure}
\centerline{\psfig{file=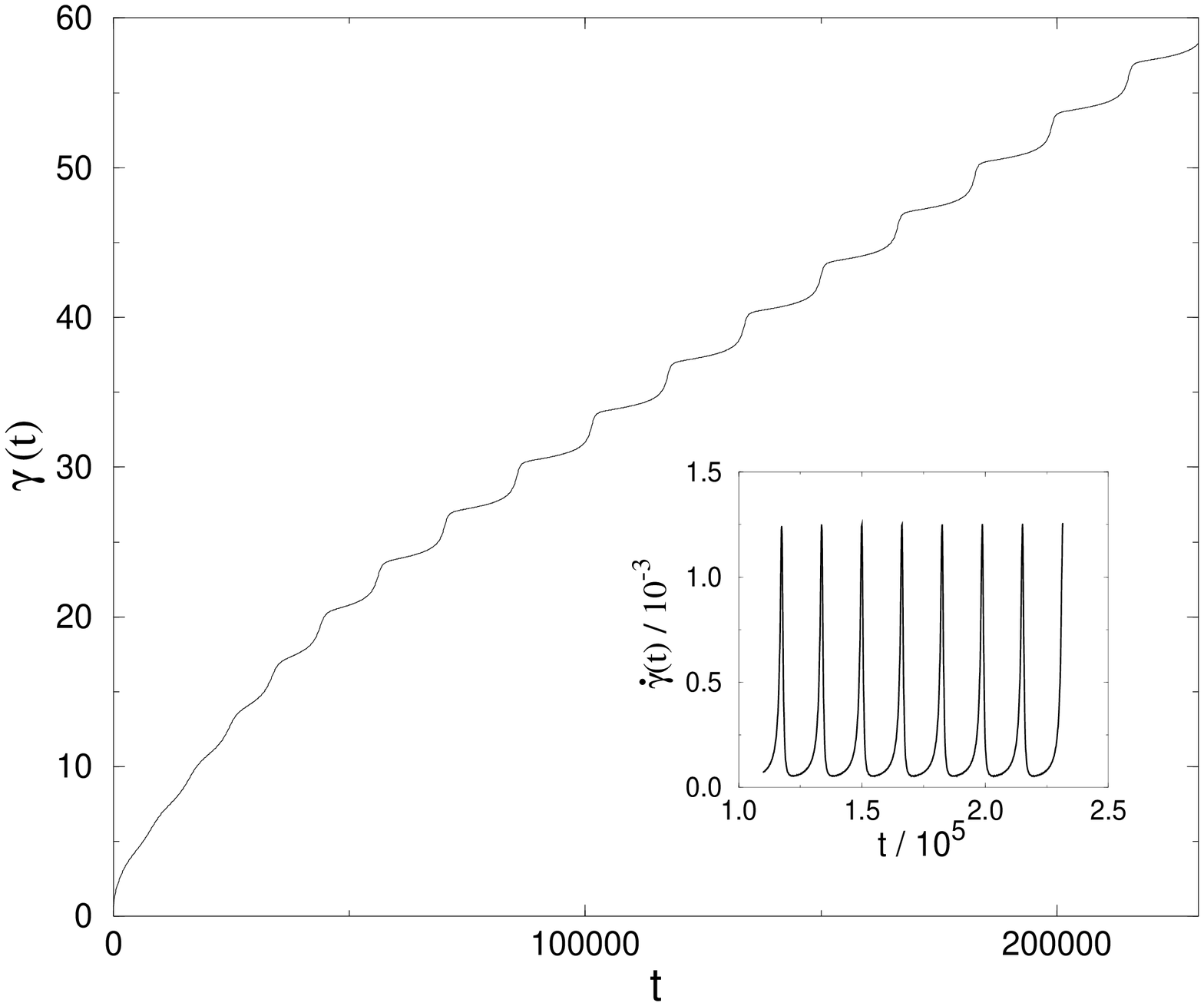,width=10cm}}
\caption{The strain $\gamma(t)$ under an imposed stress $\sigma_{0}=0.05$
for a system in which $x$ depends on the local strain according to
$x(l)=1.8$ for $l<0.4$ and $x(l)=0.8$ for $l>0.4$.
{\em (Inset)}~The strain rate $\gd(t)$ for the same data
after the transient.
}
\label{f:local_osc_eg}
\end{figure}

\begin{figure}
\centerline{\psfig{file=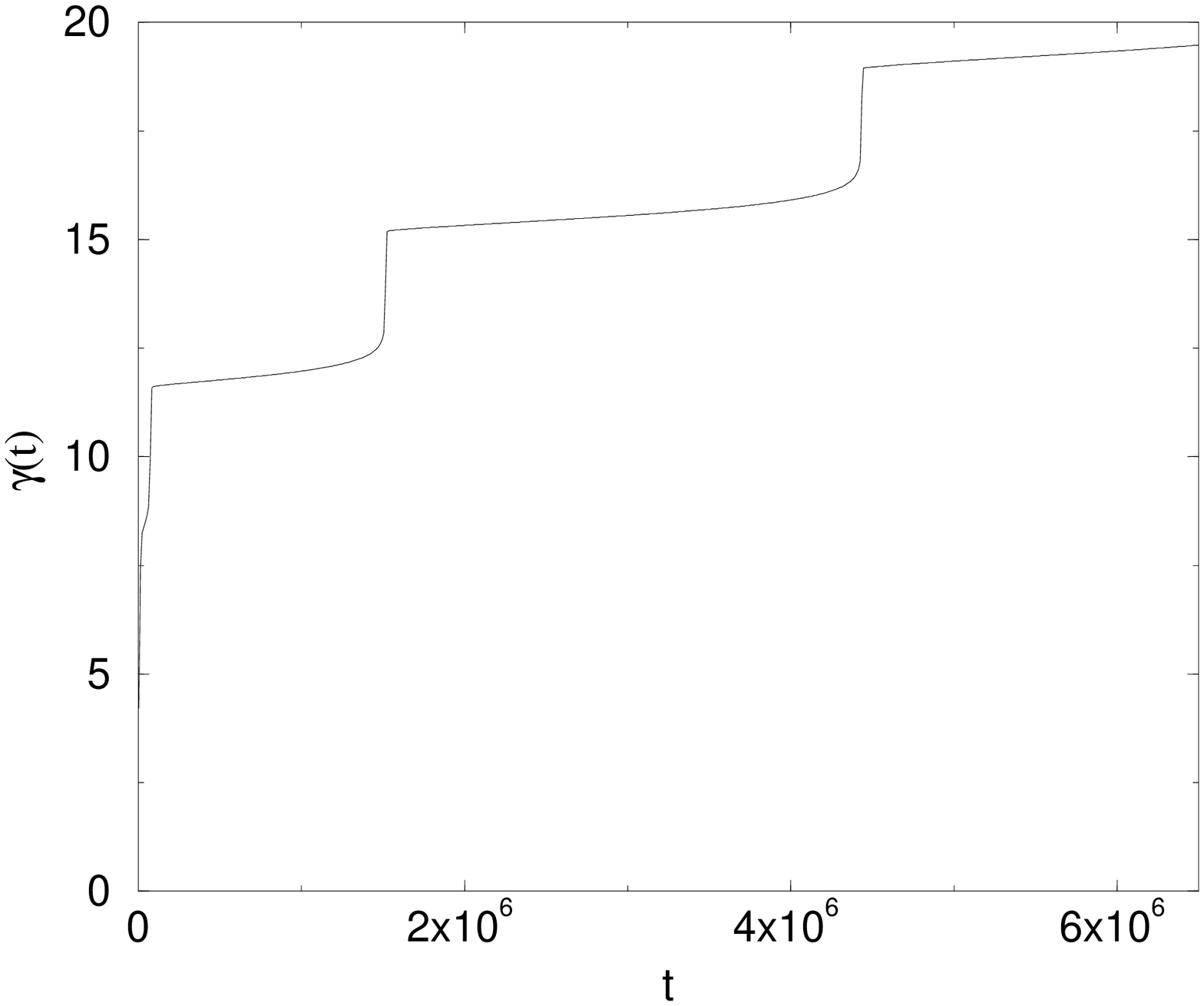,width=9cm,angle=270}}
\caption{The strain $\gamma(t)$ under an imposed stress $\sigma_{0}=0.1$
for $x(l)=1.5$ for $l<0.7$, 0.5 for $l>0.7$.
Despite the long times attained, there is no clear indication of
a single period of oscillation.
}
\label{f:local_trans_eg}
\end{figure}

An example of collective behaviour is given in Fig.~\ref{f:local_osc_eg},
which shows $\gd(t)$ for a system driven by an imposed stress $\sigma=0.05$.
As before, this plot was generated by numerically integrating the
master equation (\ref{e:master}) from an initially unstrained
state, using the procedure described in
Appendix~\ref{s:numerics}.
For this example, the system undergoes a short transient before entering
into an oscillatory regime,
in which $\gd(t)$ varies with a single period of oscillation.
There is no suggestion of a decay in the amplitude of
the oscillation in $\gd(t)$ over the largest simulation times attainable,
even when plotted against $\ln t$ (not shown),
suggesting that this is the true asymptotic behaviour.
For different choices of $x(l)$
it is not possible to identify a single period of oscillation,
as demonstrated in Fig.~\ref{f:local_trans_eg}.
It is not yet clear if this behaviour represents a distinct
regime with more than one period of oscillation,
possibly even a precursor to chaotic behaviour,
or if it is merely a long--lived transient that
eventually crosses over to either steady flow or
a single--period oscillation at later times.

The extensive simulation times means
that we have so far been unable to map out the class
of $x(l)$ that give rise to an oscillatory
or otherwise non--steady state.
Nonetheless we can make the following observation.
For all of the oscillatory cases that we have observed thus far,
$x(l)$ changes from a value in the range $1<x<2$
for small~$l$, to a value $x<1$ for larger~$l$.
Other choices of $x(l)$ may produce an oscillatory
transient, but its amplitude always seems to decay to zero in time.
It is not clear to us why only this class of $x(l)$ 
can produce a stable oscillatory regime, but it is
tempting to note that the requirement that $1<x<2$
for small $l$ is also the range of $x$ for which the
SGR model does not have a linear regime under an
imposed~$\sigma$~\cite{sgr_aging}.
That is, a significant proportion of elements will eventually
gain a finite strain, no matter how small $\sigma$ is,
and thus the variation in $x(l)$ will eventually be `felt.'

Once a suitable $x(l)$ has been found, it is possible
to move in and out of the non--steady regime by varying
the applied stress~$\sigma$.
Generally, we find that a high $\sigma$ gives rise to
steady flow, and low $\sigma$ produces oscillations;
however, the excessive simulation times needed for small $\sigma$
means that we have not been able to rule out another
crossover to steady flow as $\sigma\to0^{+}$.
Within the oscillatory regime, the mean strain rate
averaged over a period of oscillation
is generally much lower than that predicted by the flow curve.
For the examples already given, the mean strain
rate deviates from the flow curve
by a factor of 2 for the oscillations shown in Fig.~\ref{f:local_osc_eg},
and by two orders of magnitude for the 
(possibly transient) oscillations of Fig.~\ref{f:local_trans_eg}.
Again, we have been unable to fully characterise this behaviour
with the available simulation resources.
To an extent, these problems will be alleviated by considering
a simplified model considered in Sec.~\ref{s:monodisperse}.
Before turning to consider this, we provide
a qualitative description of the
oscillatory behaviour in terms of the local strains~$l$.

\subsection{Qualitative description of the evolution of the system
in the oscillatory phase}
\label{s:qual_desc}

\begin{figure}
\centerline{\psfig{file=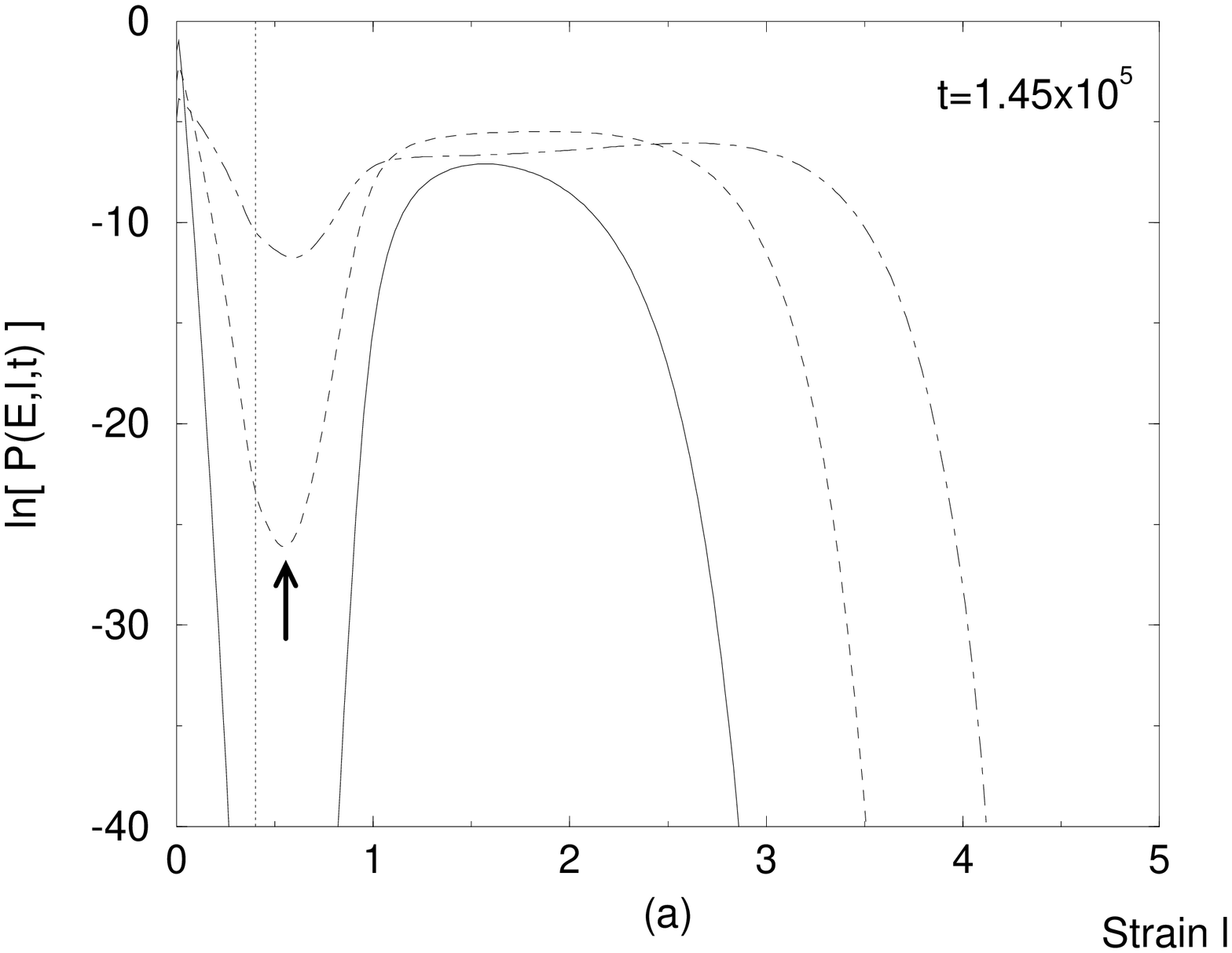,width=9cm,angle=270}}
\centerline{\psfig{file=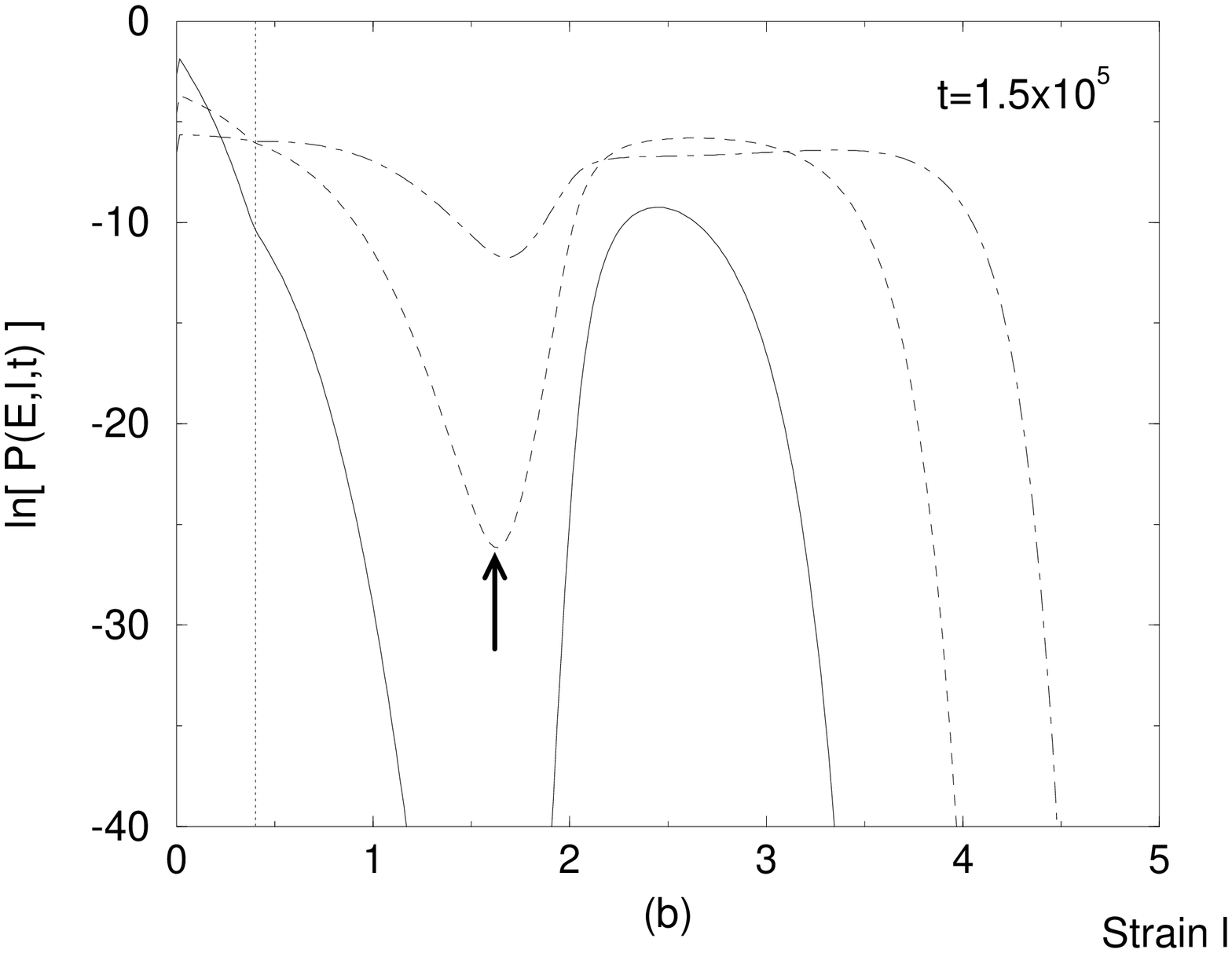,width=9cm,angle=270}}
\centerline{\psfig{file=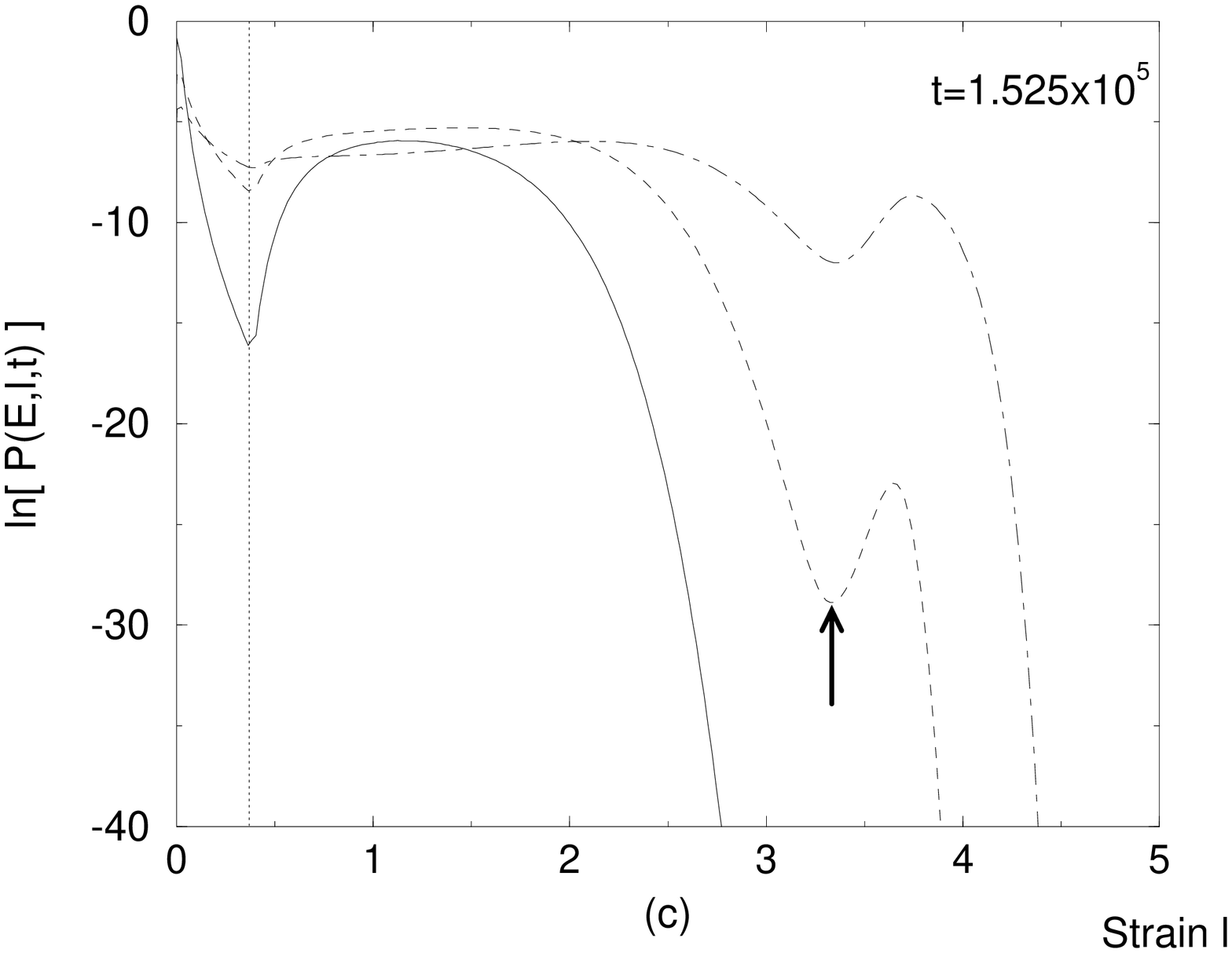,width=9cm,angle=270}}
\caption{Snapshots of $P(E,l,t)$ for the same system as in
Fig.~\ref{f:local_osc_eg} at 3 different times.
For clarity only 3 values of $E$ are shown,
namely $E=8$ (solid line), $E=10$ (dashed line) and
$E=12$ (dot--dashed line).
The chosen times correspond to just before~(a), during~(b)
and just after~(c) the point at which $\gd$ reaches it maximum value.
The period of the oscillation is approximately $\Delta t=1.6\times10^{4}$,
so that the time between (a) and (c) comprises
roughly $\frac{1}{2}$ of a single oscillation.
In each case, the $l$ at which $x(l)$ changes from 1.8 to 0.8
is represented by a vertical dotted line,
and the arrow points to the {\em same} dip in the $E=10$
distribution, which moves to the right under the
action of~$\gd$.
}
\label{f:snapshots_poly}
\end{figure}

The mechanism behind the oscillatory behaviour can be qualitatively
understood by inspecting the evolution of the $P(E,l,t)$
distribution during a single period of oscillation.
As an example, a suitable sequence of snapshots is
given in Fig.~\ref{f:snapshots_poly}.
To aid in the interpretation of these figures,
it is useful to recall some properties of the master
equation~(\ref{e:master}).
Firstly, the convective term $\gd\partial P/\partial l$ means that
any maxima or minima in $P(E,l,t)$
will move in the direction of increasing $l$
at a rate~$\gd(t)$.
These extrema will eventually disappear when large values of $l$
are reached, and the yield rate $\Gamma(E,l)$ becomes very high.
Also, there is a flux of newly--yielded elements
at $l=0$,
which have barriers $E$ weighted according to $\rho(E)$.

Since the oscillatory behaviour is by definition cyclical,
it is somewhat arbitrary where one chooses to start the sequence
of snapshots.
In Fig.~\ref{f:snapshots_poly} we have chosen times
corresponding to before, during and just after
the point when $\gd$ reaches its maximum value.
In the first snapshot,
$P(E,l,t)$ is concentrated into two regions~: a sharp peak at $l\approx0$,
and a broader peak with $l$ in the range $1<l<3$.
The first peak is `hot' in the sense that it has the higher
value of~$x(l)$, but since it has a low strain,
it does not significantly contribute to the total stress~$\sigma$.
The second peak, although highly strained, lies in a region
in which $x(l)$ is small, and therefore has a low rate of yielding.
As long as the yield rate is low, so too is the rate of stress
loss from elements that belong to this peak.
Therefore the strain rate $\gd$ will also be low,
and indeed this first example corresponds to a system in
which $\gd$ is close to its minimum value.

This state of affairs is not stable, however.
Although the yield rate of the highly strained elements
is low, it is nonetheless non--zero, and therefore so is~$\gd$.
Consequently the whole system is being convected
at a finite rate, so the elements in the
high--$l$ region are becoming more strained.
This {\em decreases} their effective energy barrier
$\eff=E-\frac{1}{2}l^{2}$,
which {\em increases} their yield rate.
But an increased yield rate means an increased~$\gd$,
which in turn increases the rate at which the elements
are becoming more strained, which increases their yield
rate yet more, and so on.
This description is that of a {\em positive feedback loop},
which causes $\gd$ to increase at ever faster rates until
all of the heavily strained elements have been depleted.
The second snapshot in Fig.~\ref{f:snapshots_poly}
show the state of the system shortly before this has happened.

At the same time that the highly--strained elements are
being depleted,
$\gd$ is close to its maximum value and consequently
$P(E,l,t)$ in the small $l$ region becomes somewhat flat,
certainly much flatter than under a small~$\gd$.
This can clearly be seen in the second and third snapshots
of Fig.~\ref{f:snapshots_poly}.
When $\gd$ again falls to lower values,
this flat part of the distribution will start to decay
as the elements within it yield.
However, the yield rate depends on~$x(l)$,
and since $x(l)$ changes to a lower value at $l=0.4$,
$P(E,l,t)$ will decay much more rapidly for $l$ just smaller
than $0.4$ than for $l$ just greater than $0.4$.
Thus a dip will occur around the point $l\approx0.4$,
which can clearly be seen in the final snapshot.
As time increases, this dip will become more pronounced
until the system can again be separated into a population
of highly--strained elements
and a second population with $l\approx0$.
This is where we began, and thus the cycle
is complete.

\section{Jamming model C: \lowercase{$x=x(l)$}, monodisperse $E$}
\label{s:monodisperse}

It is possible to more completely describe the oscillatory
regime for a simplified version of the model where every
element has the same energy barrier $E_{1}$\,.
This formally corresponds to a `monodisperse'
prior distribution $\rho(E)=\delta(E-E_{1})$,
as opposed to the exponential $\rho(E)={\rm e}^{-E}$
which has been employed in all
of the earlier sections.
The reduction in the number of degrees of freedom
that this entails 
significantly decreases the simulation times,
and therefore allows for a more thorough numerical
investigation of the model.
It is also possible to derive an analytical criterion
for the stability of the steady state,
as described below.

Furthermore, the very fact that a monodisperse system
also has an oscillatory regime clearly demonstrates
that this phenomenon is robust.
In particular, it does not require the usual SGR assumption
of an exponential tail to $\rho(E)$.
Therefore the possible existence of a yield stress,
which was so central to the
history--dependent jamming scenario
at controlled $\gd$ described in Sec.~\ref{s:global},
is not necessary.
This robustness means that the
mechanism behind the oscillations at controlled $\sigma$
may be realisable in a much broader range of models
than the restricted set considered here and,
possibly, may also occur in real materials.

The emergence of oscillations in this monodisperse
model has been separately discussed elsewhere~\cite{EPL}.
Here we supply extra details, such as the stability
analysis of the steady flow state.
For the sake of completeness, the overall behaviour of the
model will also be briefly described,
although we refer the reader to~\cite{EPL} for a fuller
discussion of many of the points raised below.

\subsection{Steady state behaviour for monodisperse $E$}

Perhaps the most immediate consequence of only allowing a
single barrier $E_{1}$ is that the system can never have
a finite yield stress.
Indeed, in the linear regime $\gd\to0^{+}$, the steady state
stress is just $\sigma\sim\gd{\em e}^{E_{1}/x(0)}$,
which vanishes with $\gd$.
There are no qualitative differences for $x<1$, $1<x<2$ {\em etc.},
as in the SGR model.
The complete lack of a yield stress means that 
a monodisperse system with $x=x(\sigma)$
will {\em not} exhibit a jamming transition, in contrast to the
situation for a $\rho(E)$ with an exponential
tail described Sec.~\ref{s:global}.

One respect in which the monodisperse $E$ model is similar
to the exponential $\rho(E)$ case is that its flow curves for $x=x(l)$
are also monotonic.
In fact, the monotonicity proof given in Appendix~\ref{s:monotonicity}
holds for arbitrary $x(l)$ and $\rho(E)$.
Thus there are no regions on the flow curve with a negative slope,
and therefore there are no ranges of control parameters
for which one might normally expect bulk shear banding to arise.
Just as in the previous section, however,
oscillatory behaviour can be realised under
an imposed constant stress.

\subsection{The oscillatory regime}

\begin{figure}
\centerline{\psfig{file=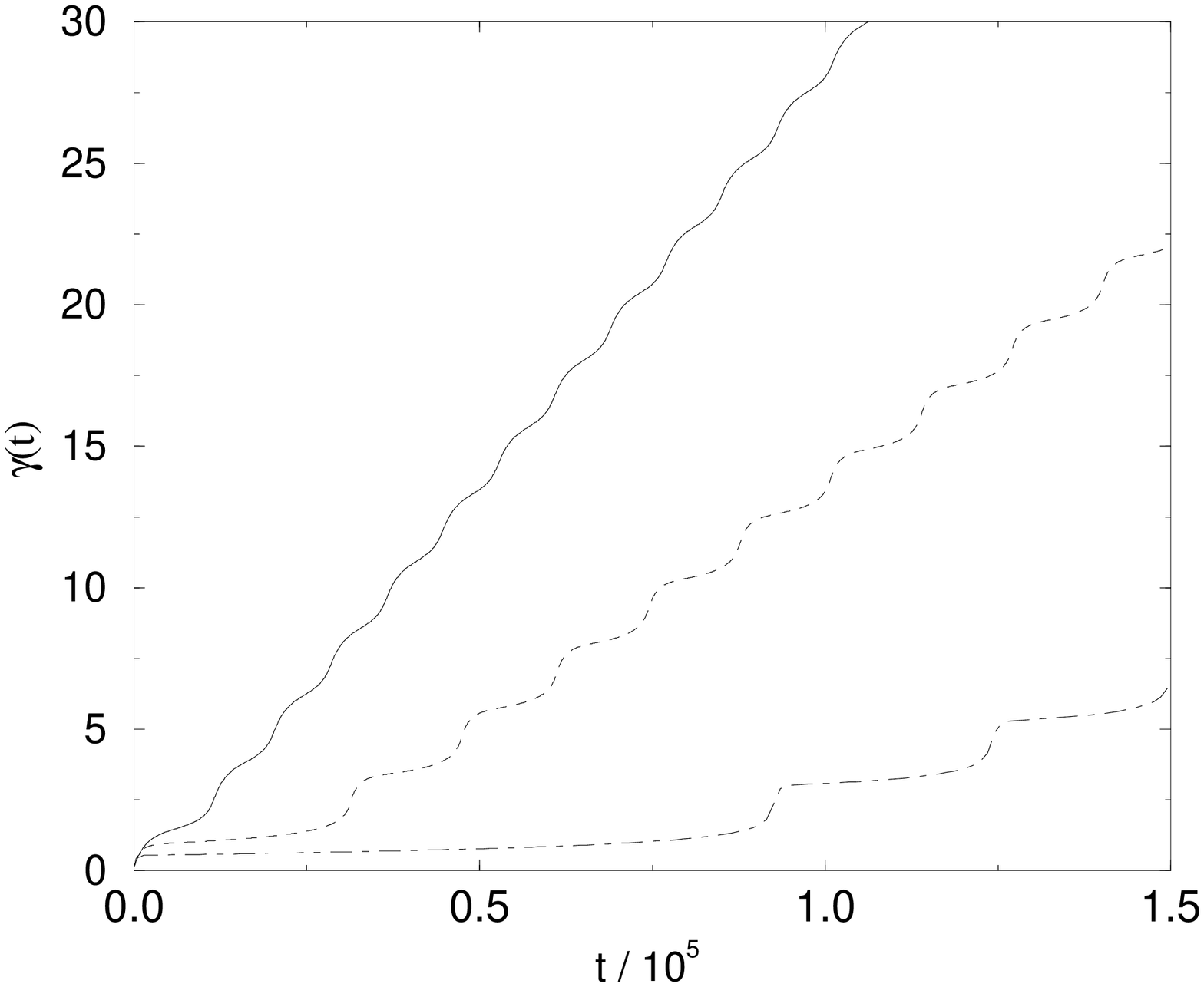,width=9cm,angle=270}}
\caption{The strain $\gamma(t)$ for a monodisperse system with
$E_{1}=5$ under an imposed step stress $\sigma=0.1$ (solid line),
0.13 (dashed line) and 0.2 (dot-dashed line) at a time $t=0$,
demonstrating a decrease in the mean $\gd$ with~$\sigma$.
Here, $x(l)=1$ for $l<0.4$ and $x=0.4$ for $l>0.4$.
The system was initially unstrained.
}
\label{f:osc_nonmon_eg}
\end{figure}

As in the polydisperse case,
the monodisperse model exhibits an oscillatory regime
under conditions of imposed stress, but not under
an imposed~$\gd$.
However, the oscillatory regime in the monodisperse model
differs from the polydisperse case in
that only single--period oscillations have so far been observed.
There is no suggestion of the more complex non--steady
behaviour hinted at earlier in Fig.~\ref{f:local_trans_eg},
for example.
Some examples of the oscillatory behaviour in the monodisperse
case are given in Fig.~\ref{f:osc_nonmon_eg}.
Here, the strain $\gamma(t)$ is shown for 3 systems with $E_{1}=5$ and
the same $x(l)$, but different imposed~$\sigma$.
Remarkably, the mean strain rate $\langle\gd\rangle$,
where the use of the angled brackets now denotes
$\gd$ averaged over a single period of oscillation,
is clearly a {\em decreasing} function of~$\sigma$,
in complete contrast to the monotonic flow curve.

\begin{figure}
\centerline{\psfig{file=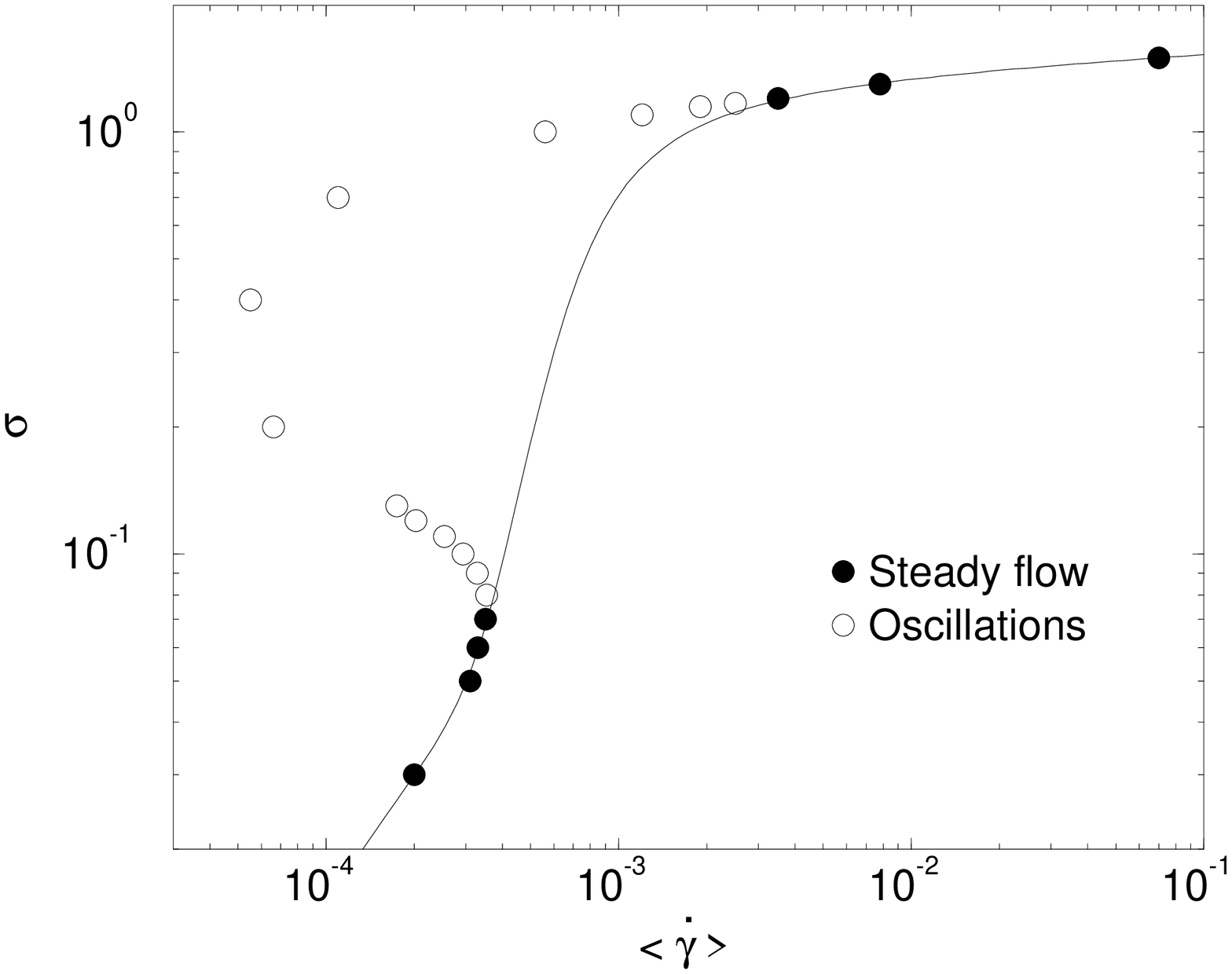,width=9cm,angle=270}}
\caption{The flow curve for the same system as in
Fig.~\ref{f:osc_nonmon_eg} under an imposed stress.
The circles are the $\gd$ as measured from the simulations,
either in the steady state (solid circles) or in the oscillatory
regime (open circles).
In the latter cases, $\gd$ was averaged over a whole number of oscillations.
The sizes of the circles are larger than the error bars.
For comparison, the theoretical flow curve for the steady
state solution $P_{\infty}(E,l)$ is plotted as a solid line.
}
\label{f:osc_flowcurve}
\end{figure}

Varying $\sigma$ over a wider range of values reveals that
steady flow is reached for either sufficiently small or sufficiently
large imposed stresses;
only for intermediate $\sigma$ are oscillations observed.
The $\langle\gd\rangle$ as read off from the simulations
are plotted against $\sigma$ in Fig.~\ref{f:osc_flowcurve},
overlayed with the steady state flow curve.
It is clear that, if the system reaches a steady state,
it falls onto the flow curve and therefore
the dependence of $\langle\gd\rangle$ with $\sigma$ is monotonic.
Within the oscillatory regime, however,
the $\langle\gd\rangle$
line deviates from the flow curve to an extent that
it even becomes non--monotonic
for a broad range of~$\sigma$.

The shape of the oscillations varies with~$\sigma$.
Close to either transition to the steady flow regime,
the oscillations are approximately sinusoidal,
indicating that there is only a single, finite frequency
of oscillation at the transition,
the amplitude of which vanishes with the onset of steady flow.
Further into the oscillatory regime,
$\gd$ can no longer be decomposed into a single harmonic,
but instead approaches a waveform in which most of the
variation in $\gd$ is compressed into
a small fraction of the total period of oscillation.
Consequently $\gamma(t)$ approaches an almost rectangular,
staircase shape.
The mechanism underlying the rapid increase in $\gd$
is the same positive feedback loop that has already been
discussed for the polydisperse case,
as can be seen by inspecting the evolution
of the $P(l)$ distribution with time~\cite{EPL}.

\begin{figure}
\centerline{\psfig{file=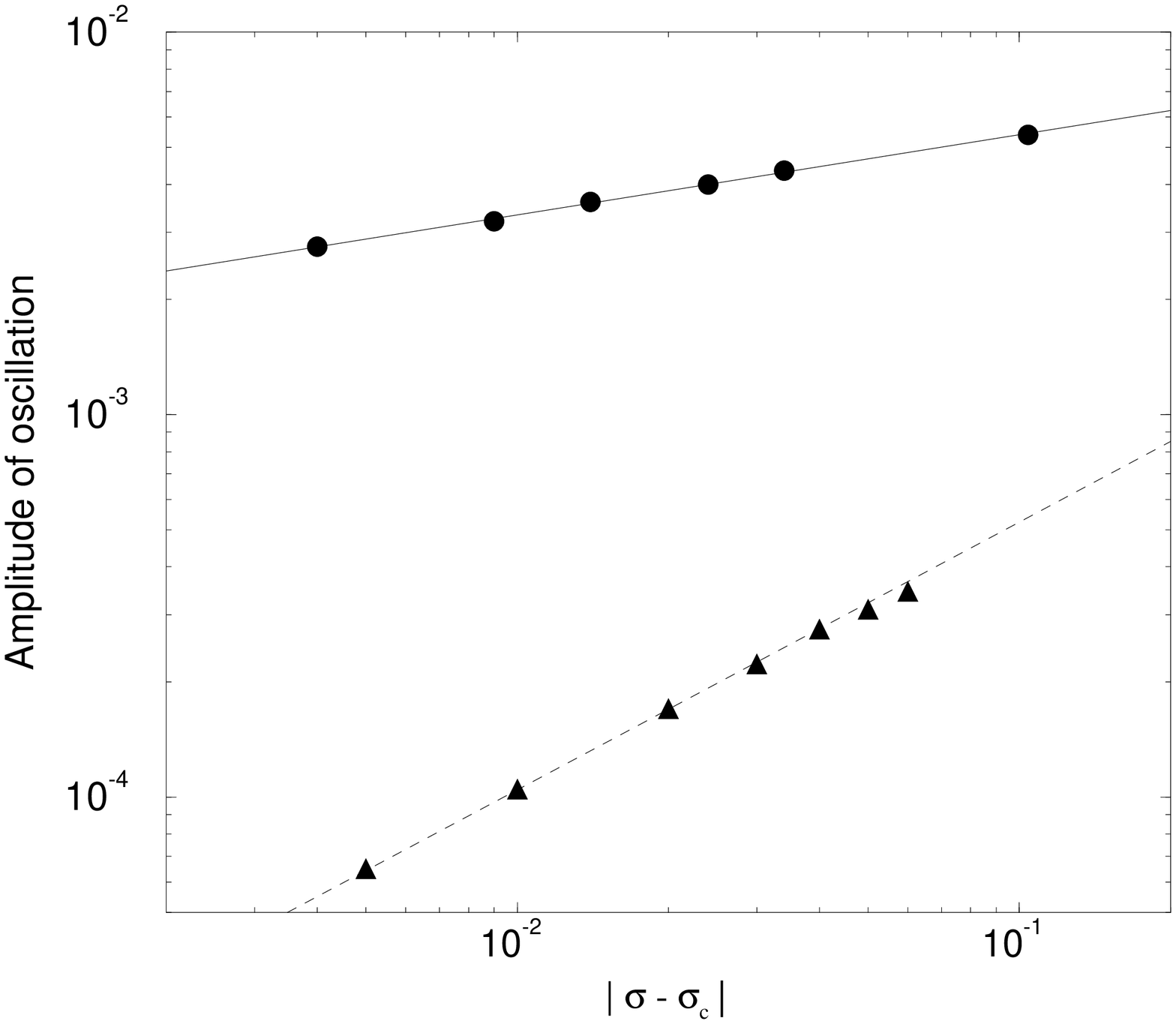,width=9cm,angle=270}}
\caption{The amplitude of the oscillation, defined
as $\frac{1}{2}|\gd_{\rm max}-\gd_{\rm min}|$, against the distance
from the transition between the steady and oscillatory regimes.
The upper data set (circles) corresponds to the transition
at $\sigma_{\rm c}=1.204$, and the lower set (triangles)
corresponds to $\sigma_{\rm c}=0.07$.
For comparison, the solid straight line has a slope of 0.21,
and the dashed line has a slope of 0.7.
}
\label{f:pow_law_div}
\end{figure}

Near to the transition points at $\sigma=\sigma_{\rm c}$,
the amplitude of the oscillation appears to vanish as
$|\sigma-\sigma_{\rm c}|^{\alpha}$ with $\alpha>0$,
as demonstrated in Fig.~\ref{f:pow_law_div}.
Data fitting suggests that the lower threshold lies
at $\sigma_{\rm c}\approx0.07$ with a value of $\alpha\approx0.7$,
and that the upper threshold is at $\sigma_{\rm c}\approx1.204$
and has a different value of $\alpha\approx0.2$.
However, the range of values given in this figure
is somewhat limited, constituting only an order of magnitude
of variation in $|\sigma-\sigma_{\rm c}|$ and an even narrower
range of amplitudes.
The reasons for this are purely technical~: close to the transition
points, the amplitude decays very slowly in time to its asymptotic value,
rendering the required simulation times impractically long.
Hence it is conceivable that the true values of $\alpha$ 
are significantly different from those found here.
In particular, a value of $\alpha=0.5$ for both cases,
as expected for a Hopf bifurcation~\cite{glendinning},
cannot be ruled out.

\begin{figure}
\centerline{\psfig{file=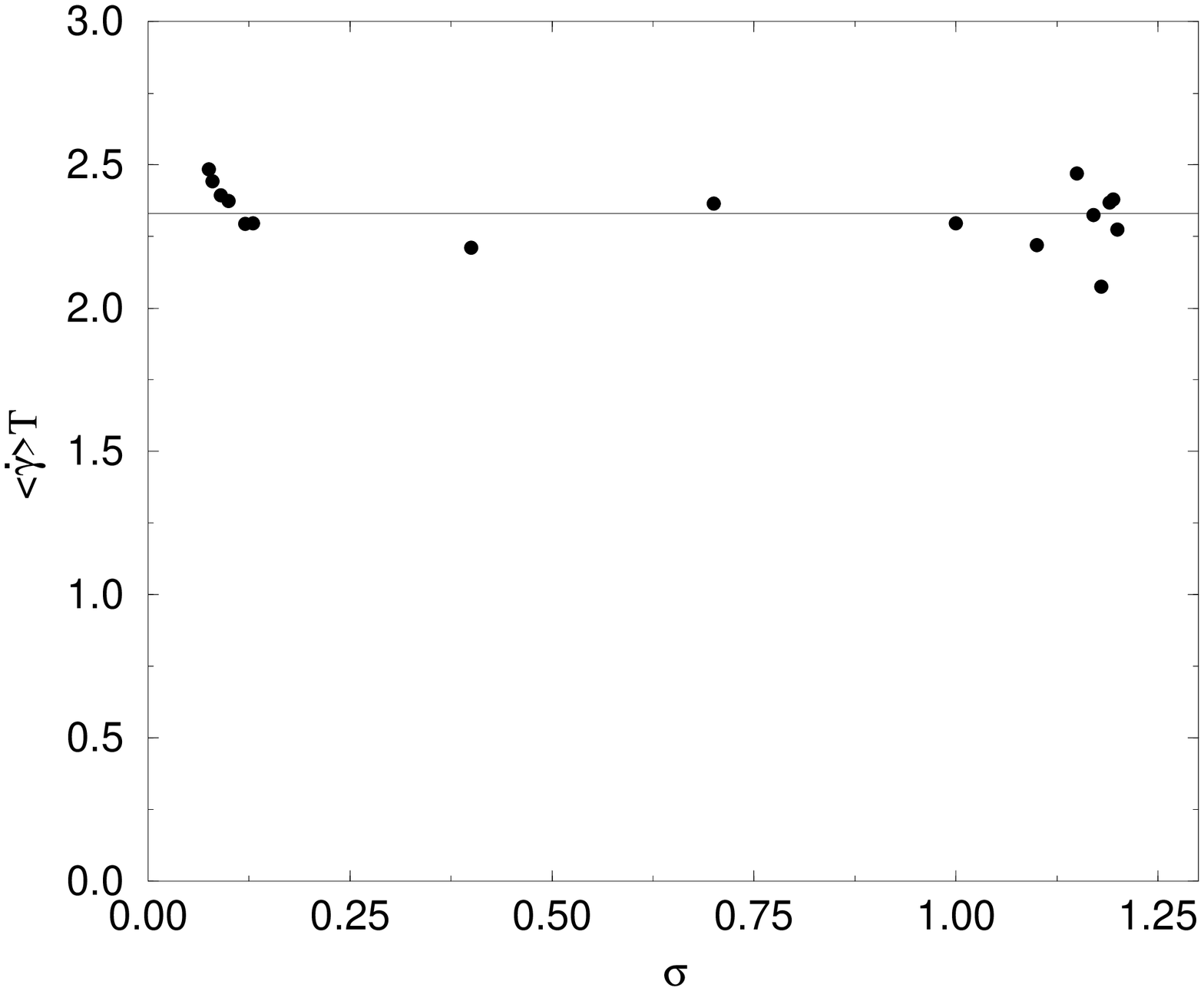,width=9cm,angle=270}}
\caption{The product of the mean strain rate $\langle\gd\rangle$ and
the period of oscillation $T$ for different stresses $\sigma$,
demonstrating that this quantity is approximately constant.
The solid horizontal line represents $\langle\gd\rangle T=2.33$.}
\label{f:constant_gd_t}
\end{figure}

Finally, plotted in Fig.~\ref{f:constant_gd_t} is the product
of the mean strain rate $\langle\gd\rangle$ and the period
of oscillation~$T$ for different values of~$\sigma$.
To first order, $\langle\gd\rangle T$ is seen
to be independent of~$\sigma$, which suggests that the oscillatory
regime can be characterised by a single
strain $l^{*}\sim\langle\gd\rangle T$.
In this case $l^{*}\approx2.3$,
which is also the typical increase in $\gamma$ during a single
cycle of oscillation.
A possible interpretation of $l^{*}$ is that
it is the amount by which the system needs to
be strained until the positive feedback loop discussed earlier
starts to dominate the system behaviour, causing it to
`reset' to the start of its cycle.
If this is correct,
then $l^{*}$ should correspond to the point at which
highly--strained elements have the same yield rate as
unstrained elements,
{\em i.e.} $\Gamma(0)\approx\Gamma(l^{*})$.
Rough calculations based on this assumption give
$l^{*}\approx\sqrt{2E_{1}[1-x(\infty)/x(0)]}$,
which for this example predicts $l^{*}\approx2.4$, in fair agreement
with the observed value.

\subsection{Analysis of the stability of the steady state}

A second advantage of the monodisperse model is that it
is possible to derive an analytical criterion for the
stability of the steady state at fixed driving shear stress~$\sigma$,
although even in this simplified case the resulting expression
is still somewhat unwieldy.
Suppose steady flow is reached with a strain rate
$\gd_{\infty}=\gd_{\infty}(\sigma)$\,.
Then the corresponding $P_{\infty}(l)$ is

\begin{equation}
P_{\infty}(l)={\cal{N}}
\exp\left[
-\;\frac{1}{\gd_{\infty}}
\int_{0}^{l}\Gamma(l')\,{\rm d}l'
\right]\quad,
\end{equation}

\noindent{}where the normalisation constant ${\cal{N}}$ is fixed
by the constraint $\int_{0}^{\infty}P_{\infty}(l)\,{\rm d}l=1$.
We now look for eigenfunctions $\{p(l),g\}$ and eigenvalues
$\{s\}$ of the linearised relaxation operator by writing

\begin{eqnarray}
P(l,t)=&P_{\infty}(l)\quad&+\quad\varepsilon\,p(l)\,{\rm e}^{st}\quad,
\label{e:stabdefp}\\
\gd(t)=&\gd_{\infty}&+\quad\varepsilon\,g\,{\rm e}^{st}\quad.
\label{e:stabdefg}
\end{eqnarray}

\noindent{}In these expressions, $g$ is a real constant,
and the real functions $p(l)$ remain to be found.
The complex constant $s$
determines the stability of the steady state~:
it is unstable precisely when ${\rm Re}(s)>0$, 
since the amplitude of the perturbation will then increase
exponentially in time.
Conversely, it is stable when ${\rm Re}(s)<0$.
The frequency of the oscillatory part
of the motion near to the steady state is
${\rm Im}(s)/2\pi$~\cite{glendinning}.

The unknown function $p(l)$ can be found by
inserting (\ref{e:stabdefp}) and (\ref{e:stabdefg})
into the master equation and neglecting terms of $O(\varepsilon^{2})$,
which gives

\begin{eqnarray}
p'(l)\gd_{\infty}+p(l)[s+\Gamma(l)]
&=&
-gP_{\infty}'(l)\\
&=&
\frac{g\Gamma(l)}{\gd_{\infty}}P_{\infty}(l)\quad.
\label{e:linearised}
\end{eqnarray}

\noindent{}This is just a first--order differential equation in~$l$
and integrates to
\begin{eqnarray}
p(l)&=&\frac{g{\cal{N}}}{\gd_{\infty}^{2}}
\left[
A+\int_{0}^{l}{\rm d}l'\,\Gamma(l')
{\rm e}^{sl'/\gd_{\infty}}
\right]\nonumber\\
&&
\times
\exp\left[-\int_{0}^{l}{\rm d}l'\,\frac{s+\Gamma(l')}{\gd_{\infty}}\right]
\label{e:ptilde}
\end{eqnarray}

\noindent{}The constant $A$ can be found by imposing
$\int_{0}^{\infty}{\rm d}l\,P(l,t)=1$, {\em i.e.}
$\int_{0}^{\infty}{\rm d}l\,p(l)=0$.

Also, since we are considering a stress--controlled system,
$\sigma=\langle l\rangle$ must remain constant and hence

\begin{equation}
\int_{0}^{\infty}{\rm d}l\,p(l)l=0\quad.
\label{e:globcons}
\end{equation}

\noindent{}This global constraint allows the value of $s$
to be specified.
Inserting (\ref{e:ptilde}) into (\ref{e:globcons}) gives,
after some manipulations, the following equation for~$s$,

\begin{eqnarray}
\int_{0}^{\infty}{\rm d}l_{1}
\int_{0}^{\infty}{\rm d}l_{2}\,&&
(l_{2}-l_{1})
P_{\infty}(l_{1})
P_{\infty}(l_{2})
\,{\rm e}^{-s(l_{1}+l_{2})/\gd_{\infty}}
\nonumber\\
&&\times
\int_{0}^{l_{2}}
{\rm d}l\,
\Gamma(l)\,
{\rm e}^{sl/\gd_{\infty}}=0
\label{e:s_eqn}
\end{eqnarray}

\noindent{}Although this equation cannot be rearranged to
give an expression for $s$ in terms of the system parameters,
as would be desired, it is still enough to show that
a purely exponential variation away from the steady flow
cannot be observed.
This follows from the observation that the first line in (\ref{e:s_eqn})
is an odd function in $l_{2}-l_{1}$, but the second line is a
strictly increasing function of $l_{2}$ when ${\rm Im}(s)=0$.
Hence the equation cannot be obeyed for a purely real~$s$,
and the transient behaviour close the the steady state
must include an oscillatory component, which is
consistent with the simulation results.

\section{Discussion}
\label{s:disc}

One of the most striking findings of this work has been the
observation of oscillations in $\gd(t)$ under a constant imposed
stress~$\sigma$ for some $x=x(l)$,
as described in Secs.~\ref{s:local} and~\ref{s:monodisperse}.
Some consequences of this phenomenon have already
been discussed in~\cite{EPL}.
Two further points will be discussed here~:
the identification of the dominant physical mechanisms,
and the relationship to so--called `stick--slip' behaviour observed
in other systems.
Both of these will be considered in turn.

\subsection{Physical picture and analogous phenomena}

In Sec.~\ref{s:qual_desc}, the mechanism behind
the oscillatory behaviour was explained
in terms of two separated populations of elements~:
a `cold' population of highly strained elements with a low~$x$,
and a `hot' population of elements with $l\approx0$ and a
high~$x$.
It was explained how the cold elements
can give rise to a positive feedback loop,
causing $\gd(t)$ to rapidly increase until
the cold elements have yielded.
At the same time, a new population of cold elements
is produced from the hot one.

Known instances of rheological instability are often
explained in terms of the spatial coexistence of subpopulations, or phases.
For instance, the temporal oscillations in viscosity observed in wormlike
micelles under an imposed stress was attributed to
a slowly fluctuating interface between
a fluid phase and shear--induced structures~\cite{hu}.
For surfactant solutions in the lyotropic lamellar phase,
it was attributed to coexisting
ordered and disordered phases~\cite{jacques}.
However, these instabilities occur in the vicinity of
non--monotonic regions of the flow curve.
Spatial instabilities, such as shear banding, can also be found
near to where the flow curve has a negative gradient~\cite{olmsted}.
By contrast, the temporal oscillations observed in our models
arise even though they are by assumption spatially homogeneous,
and the flow curves are everywhere monotonic.

This suggests that the mechanism behind the oscillatory behaviour
seen here has not yet been observed in a rheological context.
It is therefore sensible to look further afield for
analogous phenomena.
One possibility comes from mathematical biology.
The Glass--Mackey equation describes the variation in time
of the size of a population of white blood cells
in response to a hormonal control system~\cite{glassmackey}.
It has a form similar to that of a first--order differential equation,
but includes a feedback term that depends on the state variable
at an earlier time.
This {\em delay} corresponds to the
maturation time of white blood cells from stem cells.
Because of this delay,
the population size can vary in time in a non--trivial manner,
including oscillatory behaviour and chaos~\cite{glassmackey}.

That some form of time delay may also play a role
within our $x(l)$ model is clear~:
once an element becomes `cold,'
its yield rate remains very low until some time later,
when it becomes sufficiently strained that its yield rate
becomes comparable to that of `hot' elements.
Thus there is a delay between when an element first becomes cold
and when it yields,
although here the delay time is not constant but depends
on $\gamma(t)$.
Thus it is possible that the oscillatory behaviour
observed for $x=x(l)$ could be described by a simplified equation,
similar in form to the Glass--Mackey equation.
This is a particularly enticing proposition,
as the Glass--Mackey equation is capable of producing chaos,
and chaotic behaviour has also been observed in surfactant
solutions~\cite{sood1,sood2},
albeit in the signature of $\sigma(t)$ under an
imposed~$\gd$.
It is not yet clear to us if a meaningful mapping between the two
models is possible.


\subsection{Oscillatory behaviour or `stick--slip'?}

Deep into the oscillatory regime,
the waveform of $\gd(t)$ throughout a single
period of oscillation
becomes increasingly `sharp,' with
most of the variation in $\gd(t)$ occurring in just
a small fraction of the total period of oscillation.
The ratio of the maximum value of $\gd(t)$ to its minimum
has been shown to exceed two orders of magnitude,
and we see no reason to deny that greater separations
may be attained for different parameter values.

It is tempting to refer to this behaviour of $\gd(t)$
as `stick--slip' behaviour,
in which the system is `stuck' until the short duration of time
in which $\gd$ rapidly accelerates to its maximum value,
corresponding to a `slip' event.
This kind of response is also observed in
earthquakes~\cite{earthquake},
ultra--thin liquid films~\cite{thinliquids}
and granular media~\cite{granstickslip1,granstickslip2},
for example.
However, we have instead chosen to refer to the variation
in $\gd(t)$ as merely `oscillatory,'
since the underlying physics seems to be somewhat different
to the examples of stick--slip behaviour mentioned above.
In particular, the term stick--slip is usually employed
to refer to a surface friction phenomenon,
whereas the models studied in this paper have no surface,
or indeed any form of spatial definition.
They are only intended to describe the bulk behaviour of a material.
Furthermore, we only find oscillations in $\gd(t)$
under an imposed~$\sigma$, and never {\em vice versa},
whereas stick--slip seems to more usually (although not exclusively)
refer to variations in the (normal) stress under a constant
driving velocity.
Thus to avoid possible confusion,
we choose not to refer to the behaviour observed in the models
as stick--slip.

\section{Summary}
\label{s:summ}

We have introduced a range of schematic models that are
capable of exhibiting a form of jamming under
an imposed stress~$\sigma$.
The models are based on the SGR model of Sollich {\em et al.},
but differ in that the effective temperature $x$ is no longer constant,
and can instead vary with the state of the system
through either the global stress $\sigma$
or the local strain~$l$.
We have considered choices of $x$ that decrease
for increasing $\sigma$ or~$l$,
which is relevant to the study of shear thickening materials.
These models have no spatial definition,
and thus by construction cannot exhibit any form of
spatial heterogeneity.

For $x=x(\sigma)$, the flow curves can be extracted
from the known curves for constant~$x$.
Many choices of $x(\sigma)$ produce flow curves with
non--monotonic regions,
which exhibit hysteresis in $\sigma(t)$
under ramping the strain rate $\gd(t)$
first upwards and then downwards.
Furthermore, a subset of these $x(\sigma)$
also give rise to a {\em jammed} state for a range of
applied stresses,
in that the strain $\gamma(t)$ creeps logarithmically,
$\gamma(t)\propto\ln(t)$.
The criterion for this to arise is that the
curve of $x(\sigma)$ drops below the SGR yield stress
curve $\yield(x)$ when they are plotted on the same axes.
For an imposed strain rate that decays to zero at late times,
a jammed configuration was defined as one with a finite
asymptotic stress, $\sigma(t)\sim\yield>0$ as $\gd(t)\to0^{+}$
and $t\to\infty$.
It was found that whether or not a jammed configuration was reached
depends on the entire strain history of the system,
a situation that
was referred to as {\em history dependent jamming}.

For $x=x(l)$, the flow curves are always monotonic,
and steady flow is always reached under a
constant imposed strain rate $\gamma(t)\sim\gd t$.
However, for a range of imposed stresses
and some choices of $x(l)$, steady flow is not realised.
Numerical integration of the master equation demonstrated
that $\gd(t)$ oscillated around a well--defined mean
with a single period of oscillation.
The possibility of more complex
non--steady behaviour in some regions of parameter
space could not be ruled out.
A similar oscillatory behaviour occurs
with a simpler model in which every element
has the same energy barrier~\cite{EPL},
which suggests that this phenomenon is robust.
Finally, we discussed the relationship between this
oscillatory behaviour and that observed in experiments,
and considered analogous phenomena from fields
outside of rheology.

\section*{Acknowledgements}
\label{s:ack}

The authors would like to thank Suzanne Fielding for
stimulating discussions concerning this work.
AA also wishes to thank the University of Edinburgh
for the hospitality that allowed this work to start.
DAH was funded by EPSRC(UK) grant no. GR/M09674.

\appendix
\section{Simulation details}
\label{s:numerics}

Direct numerical integration of the master equation (\ref{e:master})
has proven to be unstable with respect to discretisation errors.
Instead, the results in this paper were generated from the
transformed equation

\begin{eqnarray}
\frac{\partial P(E,\Delta l,t)}{\partial t}
&=&
\,-\,\Gamma(\Delta l+\gamma(t))\,P(E,\Delta l,t)\nonumber \\
&&\mbox{}+\hop(t)\,\delta(\Delta l+\gamma(t))\,\rho(E)\quad,
\label{e:dl_master}
\end{eqnarray}

\noindent{}where $\Delta l=l-\gamma(t)$.
This removes the convective term and
dramatically improves numerical stability.
The discrete probability distribution
$P_{ij}=P(i\delta E,j\delta l)$ was defined on a rectangular
mesh of points $\{ij\}$ and evolved according to (\ref{e:dl_master})
by using a straightforward Euler method.
A further refinement was to average both the evolution
equation (\ref{e:dl_master}) and the initial conditions
over the ranges $(E,E+\delta E)$ and
($\Delta l,\Delta l+\delta l$).
This significantly reduces the number of mesh
points required for the simulations to properly converge,
without unduly increasing the algorithmic complexity.
The delta function was treated as a triangle of base width
$2\delta l$ and height $1/\delta l$,
but any sufficiently narrow function gave the same results.

Two classes of initial conditions were employed,
but the long--time behaviour of the system was found
to be identical in both cases.
Only the short term behaviour varied, and then only in
a non--essential manner.
For the sake of completeness, the initial conditions were~:
(i)~A `quench' $P_{0}(E,l)=\delta(l){\rm e}^{-E}$, which corresponds
to the unstrained equilibrium at $x=\infty$, or
(ii)~$P_{0}(E,l)=\delta(l)\frac{1}{E_{0}}{\rm e}^{-E/E_{0}}$
with $E_{0}=[1-1/x(l=0,\sigma=0)]^{-1}$, which corresponds to
an unstrained system that has been allowed to reached equilibrium.
Note that this second initial state is only defined
for $x(l=0,\sigma=0)>1$~\cite{trap}.

The strain $\gamma(t)$ is only known {\em a priori} for a
strain--controlled system.
When it is rather a known stress $\sigma(t)$ that is applied,
$\gamma(t)$ must be chosen at every time step
so that the actual stress does not differ
from $\sigma(t)$ by more than a tolerance parameter~$\varepsilon\ll1$.
To ensure that this condition was satisfied in our simulations,
a series of trial strain rates $\gd^{(1)},\gd^{(2)},\gd^{(3)}\ldots$
were generated and tested on a replica mesh $P^{*}_{ij}\,$.
The $P_{ij}$ were not updated until a suitable $\gd$ had been found.

The trial values $\gd^{(i)}$ were generated as follows.
For a continuous time variable,
$\gd(t)=\langle l\,\Gamma(l)\rangle_{P(t)}+\dot{\sigma}(t)$ exactly,
as seen by multiplying the master equation (\ref{e:master})
by $l$ and integrating over all $E$ and~$l$.
This is therefore the sensible choice for $\gd^{(1)}$.
However, integrating over a finite time step $\delta t$
inevitably introduces errors of $O(\delta t)$,
and so the integrated stress will differ from the required value by
some small amount $\delta\sigma$.
Reintegrating with $\gd^{(2)}=\gd^{(1)}-\delta\sigma$ will therefore
reduce the error to a smaller amount $O(\delta t^{2})$.
This procedure can be repeated to generate a series of
successively better estimates $\gd^{(3)}$, $\gd^{(4)}$ {\em etc.}
For our choice of $\varepsilon=10^{-6}$, we have found that typically
3---5 such trials are needed at every time step.

\section{Monotonicity of the flow curves}
\label{s:monotonicity}

The purpose of this appendix is to show that the flow curves
are monotonic for any yield rate $\Gamma(l)$ that depends only
on the local strain~$l$.
This includes the thermally activated $\Gamma(l)$ with $x=x(l)$,
possibly constant,
but not when $x$ also depends on the stress~$\sigma$.
It also applies for an arbitrary prior barrier
distribution $\rho(E)$.

The steady state solution $P_{\infty}(E,l)$ has already
been given in (\ref{e:st3}).
The asymptotic yield rate
$\hop_{\infty}\equiv\lim_{t\rightarrow\infty}\hop(t)$
is fixed by ensuring that this expression
normalises to unity,

\begin{equation}
\hop_{\infty}^{-1}=\frac{1}{\dot{\gamma}}
\int{\rm d}E
\int{\rm d}l
\:\rho(E)
\,{\rm e}^{-f(\sigma,l)/\dot{\gamma}}
\label{e:mono_hop}
\end{equation}

\noindent{}where we have introduced the short--hand notation
$f(\sigma,l)=\int_{0}^{l}\Gamma(\sigma,l')\,{\rm d}l'$.
The stress is $\sigma=\langle l\rangle_{\infty}$,
where the angled brackets `$\langle\rangle_{\infty}$'
denote the average over $P_{\infty}(E,l)$.
If $\Gamma$ depends on~$\sigma$,
then a $\sigma$ must be chosen that is consistent with~(\ref{e:st3}).
In general there can be more than one suitable~$\sigma$.

First consider the case $\Gamma=\Gamma(l)$, so $f=f(l)$.
Then by differentiating $\langle l\rangle_{\infty}$ and using
(\ref{e:st3}),

\begin{equation}
\frac{\partial\sigma}{\partial\dot{\gamma}}
=
\frac{\sigma}{\hop_{\infty}}
\frac{\partial\hop_{\infty}}{\partial\dot{\gamma}}
-\frac{\sigma}{\dot{\gamma}}
+\frac{1}{\dot{\gamma}^{2}}
\left\langle
lf(l)
\right\rangle_{\infty}\quad.
\label{e:mono_local}
\end{equation}

\noindent{}Similarly, the $\hop_{\infty}$ equation~(\ref{e:mono_hop})
can be differentiated to give

\begin{equation}
\frac{1}{\hop_{\infty}}
\frac{\partial\hop_{\infty}}{\partial\dot{\gamma}}=
\frac{1}{\dot{\gamma}}
-\frac{1}{\dot{\gamma}^{2}}
\left\langle
f(l)
\right\rangle_{\infty}
\end{equation}

\noindent{}Combining these two expressions produces

\begin{eqnarray}
\dot{\gamma}^{2}\,
\frac{\partial\sigma}{\partial\dot{\gamma}}
&=&
\langle lf(l)\rangle_{\infty} - \sigma\langle f(l)\rangle_{\infty}
\\
&=&
\left\langle\:
(l-\sigma)
f(l)
\:\right\rangle_{\infty}
\\
&=&
\left\langle\:
(l-\sigma)\:
[f(l)-f(\sigma)]
\:\right\rangle_{\infty}
\label{e:mono_ans}
\end{eqnarray}

\noindent{}The final line in this equation is valid as
$f(\sigma)$ is a constant,
so $\langle (l-\sigma)f(\sigma)\rangle_{\infty}
=f(\sigma)\langle l-\sigma\rangle_{\infty}=0$.
Since $f(l)$ is an increasing function of~$l$,
the quantity in the angled brackets in (\ref{e:mono_ans})
is positive both for $l<\sigma$ and for $l>\sigma$,
and vanishes smoothly at $l=\sigma$.
Thus the left hand side must be positive,
{\em i.e.} $\sigma$ always increases with $\gd$, as claimed.

The same conclusion does not hold when $\Gamma=\Gamma(\sigma,l)$.
Since $\sigma$ depends on $\gd$,
differentiating the steady state solution now
gives rise to an extra term on the right hand side of (\ref{e:mono_local}),

\begin{eqnarray}
\frac{\partial\sigma}{\partial\dot{\gamma}}
&=&
\frac{\sigma}{\hop_{\infty}}
\frac{\partial\hop_{\infty}}{\partial\dot{\gamma}}
-\frac{\sigma}{\dot{\gamma}}
+\frac{1}{\dot{\gamma}^{2}}
\left\langle
lf(\sigma,l)
\right\rangle_{\infty}\nonumber\\
&&-\frac{1}{\gd}\,\langle l\,g(\sigma,l)\rangle_{\infty}\,
\frac{\partial\sigma}{\partial\gd}
\label{e:mono_global}
\end{eqnarray}

\noindent{}using
$g(\sigma,l)=\int_{0}^{l}[\partial\Gamma(\sigma,l')/\partial\sigma]{\rm d}l'$.
Proceeding as before,

\begin{equation}
\dot{\gamma}^{2}\,
\frac{\partial\sigma}{\partial\dot{\gamma}}
=
\frac{
\left\langle\:
(l-\sigma)
f(\sigma,l)
\:\right\rangle_{\infty}
}
{
1+\displaystyle{
\frac{1}{\gd}
\langle l\,g(\sigma,l)\rangle_{\infty}
}
}
\end{equation}

\noindent{}Thus it is now possible for the gradient to diverge
if $\langle l\,g(\sigma,l)\rangle$ is sufficiently negative,
{\em i.e.} if $\Gamma(\sigma,l)$ decreases sufficiently rapidly
with $\sigma$.
For the particular case of
$\Gamma(\sigma,l)={\rm e}^{-(E-\frac{1}{2}l^{2})/x(\sigma)}$,
the gradient of the flow curve diverges at any point such that

\begin{equation}
x'(\sigma)
=
\frac{-x^{2}(\sigma)\,\gd}
{
\left\langle
l
\int_{0}^{l}(E-\frac{1}{2}l'^{2}){\rm e}^{-(E-\frac{1}{2}l'^{2})/x}
\,{\rm d}l'
\right\rangle_{\infty}
}
\quad.
\end{equation}

\noindent{}We can see no obvious physical interpretation of this
mathematical criterion.



\end{multicols}

\end{document}